\documentclass[
pra, 
aps,
twocolumn,
superscriptaddress,
longbibliography,
floatfix,
notitlepage]{revtex4-2}

\usepackage{amsmath,amsfonts,amssymb,amsthm,bm,times,dcolumn}
\usepackage{graphicx}
\usepackage{mathtools}
\usepackage{booktabs}
\usepackage{array}
\usepackage{float}
\usepackage{soul}
\usepackage[normalem]{ulem}
\usepackage{longtable}
\usepackage[utf8]{inputenc}
\usepackage[english]{babel}
\usepackage[T1]{fontenc}
\usepackage{lmodern}
\usepackage{microtype}
\usepackage{gensymb}
\usepackage{physics}
\usepackage{xcolor}
\usepackage{hyperref} 

\newcommand{\WN}{\left|W_N\right\rangle}
\newcommand{\WbarN}{\left|\overline{W}_N\right\rangle}
\newcommand{\WL}{\left|W_N^{L}\right\rangle}
\newcommand{\WLb}{\bigl|\overline{W_{N}^{L}}\bigr\rangle}
\newcommand{\Cl}{C_{\ell_1}}
\newcommand{\CVB}{\mathcal{C}_{VB}}
\newcommand{\CBB}{\mathcal{C}_{BB}}
\newcommand{\CW}{\mathcal{C}_{W}}

\begin{document}
	
\preprint{APS/123-QED}


\title{Geometry versus excitation sector in the decoherence of \\ asymmetric $N$-qubit $W$ states}

\author{Sougata Bhattacharyya}
\email{msc2503121011@iiti.ac.in}
\affiliation{Department of Astronomy, Astrophysics and Space Engineering, Indian Institute of Technology Indore, Indore 453552, India}

\author{Sovik Roy}
\email{s.roy2.tmsl@ticollege.org}
\affiliation{Department of Mathematics, Techno Main Salt Lake, Techno India Group, EM-4/1, Sector V, Kolkata 700091, India}

\author{Fatih Ozaydin}
\email{mansursah@gmail.com}
\affiliation{Data Science \& AI Major, Tokyo International University, 4-42-31 Higashi-Ikebukuro, Toshima-ku, Tokyo 170-0013, Japan}
\affiliation{Institute for International Strategy and Emerging Technologies, Tokyo International University, 4-42-31 Higashi-Ikebukuro, Toshima-ku, Tokyo 170-0013, Japan}
\affiliation{Nanoelectronics Research Center, Kosuyolu Mah., Lambaci Sok., Kosuyolu Sit., No.~9E/3, Kadikoy, Istanbul 34718, T\"urkiye}

\date{\today}


\begin{abstract}
	We investigate how network geometry and excitation sector separately control pairwise entanglement decay in asymmetric multipartite $W$ states. To disentangle these effects, we introduce an analytically tractable $N$-qubit generalization of the asymmetric Lohmayer geometry and its complementary-excitation partner, yielding inequivalent vertex-base (VB) and base-base (BB) pair classes that can be compared directly with symmetric $W$-state references. We derive closed-form concurrence dynamics under representative one-sided noise models and find that, within either excitation sector, the VB concurrence has exactly the same noise dependence as the corresponding symmetric reference, preserving a noise-independent proportional advantage wherever both remain entangled. The amplitude-damping reordering previously identified for the three-qubit Lohmayer state is therefore a cross-sector effect rather than an intrinsic fragility of the VB geometry. In contrast, the BB pair exhibits a genuine same-sector structural fragility, with lower entanglement-sudden-death thresholds than the VB pair under depolarizing noise and, in the $(N-1)$-excitation sector, under amplitude damping. The results establish network geometry, excitation sector, and noise symmetry as distinct ingredients governing pairwise entanglement robustness in asymmetric quantum networks.
\end{abstract}

\maketitle

\section{Introduction}
\label{sec:intro}

Entanglement shared among more than two parties is the operational
resource behind most quantum technologies, and its classification is
already nontrivial at the smallest nontrivial system
size~\cite{horodecki2009quantum, nielsen2010quantum}. Three-qubit pure states split into two classes that cannot be interconverted by stochastic
local operations and classical communication (SLOCC), represented by
the $|GHZ\rangle$ and $|W\rangle$ states~\cite{dur2000three}, a
classification later extended to mixed three-qubit states and to four
qubits~\cite{acin2001classification, verstraete2002four}. The symmetric
$N$-qubit $|W\rangle$ state distributes a single excitation uniformly
over all $N$ qubits and is the single-excitation member of the Dicke
family~\cite{dicke1954coherence}. Its defining operational feature is
persistence under particle loss: whereas the two-qubit correlations of
a $|GHZ\rangle$ state vanish as soon as any single qubit is traced
out, $|W\rangle$-class states retain pairwise entanglement after the
loss of one or more qubits~\cite{dur2001multipartite,neven2018entanglement}.
This is precisely the regime relevant to realistic devices, in which
local imperfections and partial access are the rule rather than the
exception.

This robustness has made $|W\rangle$-class resources attractive across
a broad range of tasks. They serve as nodes and links of entanglement
distribution architectures, quantum repeaters and the projected
quantum internet~\cite{kimble2008quantum, briegel1998quantum, azuma2015all, cacciapuoti2019quantum, acin2007entanglement,bayrakci2022quantum}; suitably chosen $W$-type states support perfect teleportation and superdense coding~\cite{bennett1993teleporting, agrawal2002probabilistic, agrawal2006perfect, li2016generating}; and multipartite entanglement underpins secret sharing and threshold cryptographic protocols~\cite{hillery1999quantum, tokunaga2005threshold}. In quantum metrology, multipartite entanglement is the resource that
pushes phase estimation beyond the standard quantum
limit~\cite{toth2012multipartite, pezze2018quantum, degen2017quantum, huang2024entanglement,ozaydin2015quantum,erol2014analysis}, 
and the fragility of this advantage under noise is well documented by exploring how standard decoherence channels degrade the quantum Fisher
information of $|W\rangle$ states~\cite{ozaydin2014phase}.
More recently, multipartite entangled states have been employed as
quantum fuel and as working media in quantum thermal machines, where
collective and coherent resources modify work extraction, efficiency
and thermalization
rates~\cite{hardal2015superradiant, turkpencce2016quantum, altintas2014, niedenzu2018quantum, manatuly2019collectively, tuncer2019work}, in particular with $|W\rangle$ states in focus~\cite{dag2019temperature, ozaydin2024engineering}.

The price of this versatility is that $|W\rangle$ states are hard to
prepare. Unlike $|GHZ\rangle$ and graph states, which follow from
short Clifford circuits, $|W\rangle$ states require dedicated
protocols, and a substantial body of work has been devoted to them:
probabilistic optical fusion of smaller $W$ states and its
optimizations~\cite{ozdemir2011optical,bugu2013enhancing,yesilyurt2013optical,ozaydin2014fusing},
fusion in cavity-QED, quantum-dot and nitrogen-vacancy
platforms~\cite{zang2015generating,han2015effective,han2017effective},
deterministic expansion and size-doubling
schemes~\cite{yesilyurt2015optical,yesilyurt2016deterministic,zang2016deterministic,ozaydin2021deterministic},
quantum-eraser-assisted generation~\cite{kim2020efficient},
preparation via Pauli spin blockade in double quantum
dots~\cite{bugu2020preparing}, spin-torque
driving~\cite{sharma2020generation}, and collision-model
approaches~\cite{ccakmak2019robust}. Parallel
efforts address the wider Dicke family, including universal
transformation gates and resource-efficient preparation and expansion
circuits, also under restricted qubit
access~\cite{kobayashi2014universal,bartschi2019deterministic, mukherjee2020preparing, aktar2022divide, thapa2025expanding, vu2026intelligent}.
Experimental realizations of $W$ and Dicke states have been reported
in photonic, trapped-ion and atomic-ensemble
systems~\cite{kiesel2007experimental,Haffner2005,chiuri2012experimental,lucke2014detecting},
and the hardware platforms on which such states must survive --
trapped ions, superconducting circuits and spin qubits -- are
precisely those characterized by the noisy intermediate-scale quantum
(NISQ) regime~\cite{haffner2008quantum,kjaergaard2020superconducting,preskill2018quantum}.

Almost all of this literature concerns \emph{symmetric} $W$ states.
Asymmetry, however, is neither exotic nor merely a defect. It arises
generically from imperfect or partial state preparation, from
deliberately engineered network topologies in which nodes play
inequivalent roles, and from the failure branches of probabilistic
fusion and expansion protocols, which frequently return recyclable
$W$-like states with unequal
amplitudes~\cite{ozdemir2011optical,thapa2025expanding}. Asymmetry can
also be a resource in its own right: perfect teleportation and perfect
superdense coding require $W$-like states with specifically
\emph{unequal} superposition amplitudes rather than the symmetric
$|W\rangle$ state~\cite{agrawal2006perfect,li2016generating}. The
three-qubit state introduced by Lohmayer \textit{et al.}~\cite{lohmayer2006entangled} is the minimal and cleanest instance
of such a structure. An unequal distribution of excitation amplitude
produces inequivalent bipartite reductions, and, because the total
pairwise entanglement that a qubit can share is
constrained~\cite{coffman2000distributed}, it also produces a nontrivial
allocation of that entanglement among the links of the network. A
comparable asymmetry is largely unexplored in quantum thermodynamics,
where multipartite fuel and working states are almost invariably taken
to be permutation
symmetric~\cite{hardal2015superradiant,turkpencce2016quantum,manatuly2019collectively,dag2019temperature,ozaydin2024engineering},
even though directional, one-way correlations are known to alter the
thermodynamic behaviour of multipartite open
systems~\cite{zhang2023effects}. Characterizing how asymmetric
$W$-class resources decohere is therefore a prerequisite for assessing
whether they can offer any advantage in such settings.

A convenient way to organize this question is geometric. Assigning to
each pair of qubits its pairwise concurrence turns a multiqubit state
into a weighted graph, so that a permutation-symmetric state
corresponds to a complete graph with a single edge weight, while an
asymmetric state acquires a distinguished node and two or more edge
classes. For the three-qubit Lohmayer state this graph is an isosceles
triangle: a \emph{vertex} qubit is joined to each of two \emph{base}
qubits by a higher-concurrence bond, while the two base qubits are
joined to each other by a weaker one. Related structural and
topological readings of multipartite entanglement, in terms of link
splitting, Borromean and Hopf-type linking and residual coherence,
have been developed for tripartite and Dicke
states~\cite{aravind1997borromean, kauffman2002quantum, sugita2007borromean,bhattacharyya2026symmetric,bhattacharyya2026entanglement}.
The geometric picture immediately suggests a question that is natural
for any entanglement network: does concentrating pairwise entanglement
into a higher-concurrence link make that link more robust, or does the
concentration itself introduce a new form of fragility?

Answering this requires following the pairwise entanglement through
realistic decoherence~\cite{zurek2003decoherence, breuer2007the}. A distinctive
feature of open-system entanglement dynamics is that entanglement can
vanish completely at a finite noise strength, long before coherence
itself is exhausted -- entanglement sudden death (ESD) -- as
established for two-qubit systems and observed
experimentally~\cite{yu2004finite, yu2009sudden, almeida2007environment, huang2007necessary}, and
subsequently examined for tripartite and multipartite
resources~\cite{weinstein2009tripartite, carvalho2004decoherence, zyczkowski2001dynamics, lopez2008dynamics, ma2012entanglement, bellomo2007non, gao2008entanglement, siomau2010entanglement, espoukeh2015lower}.
We therefore work with the four standard single-qubit channels, each
of which probes a physically distinct mechanism. Phase damping (PD)
models pure dephasing, i.e.\ $T_{2}$ processes that destroy coherence
without energy exchange, and is the dominant error in many solid-state
spin systems. Amplitude damping (AD) models spontaneous emission and
energy relaxation ($T_{1}$) into a zero-temperature environment, and is
directional in excitation number, which makes it the natural probe of
the excitation content of a state. Depolarization (DP) is isotropic and
carries no preferred basis or excitation direction; it is the standard
worst-case benchmark noise and, for exactly that reason, a useful
control against which basis-dependent effects can be identified.
Finally, the generalized amplitude damping channel (GAD) describes a
reservoir at finite temperature, interpolating between relaxation and
excitation as its thermal weight is
varied~\cite{srikanth2008squeezed, bennett1996mixed}. GAD is also the channel that connects the present analysis to thermodynamic settings, where the
environment is by construction a finite-temperature bath and the
balance between absorption and emission determines the performance of
the machine~\cite{niedenzu2018quantum, manatuly2019collectively, dag2019temperature, ozaydin2024engineering}.

At $N=3$, our recent work~\cite{bhattacharyya2026super} compared the
two-excitation Lohmayer state with the conventional single-excitation
symmetric $|W_{3}\rangle$ state and reported a channel-dependent
reordering of the initial hierarchy $\CVB>\CW>\CBB$ under amplitude
damping, termed the \emph{Super-Link Fragility Effect}. Because
amplitude damping is sensitive to excitation number, however, that
comparison necessarily mixes the effects of network geometry and
excitation sector. A same-sector comparison is possible by including
the globally bit-flipped symmetric partner
$|\overline{W}_{3}\rangle$, but the three-qubit setting alone does not
reveal how the resulting distinction scales with system size. What is
needed is a family of states in which either the network geometry or
the excitation sector can be held fixed while the other is varied, for
arbitrary $N$.

Accordingly, we construct an $N$-qubit generalization of the Lohmayer
geometry, in which a single vertex qubit is distinguished from $(N-1)$
equivalent base qubits, together with its bit-flipped partner
occupying the complementary excitation sector. Together with the
symmetric family $\{\WN,\WbarN\}$, this gives four states spanning two
network geometries and two excitation sectors, allowing
symmetric-versus-asymmetric comparisons to be made within either the
single-excitation or the $(N-1)$-excitation sector, as well as
explicitly across the two sectors. We derive closed-form pairwise
concurrences for all four states under the four channels introduced
above. Throughout, we adopt a one-sided local-noise convention: in
every bipartite subsystem, the channel acts on one qubit while its
partner is left untouched. For a vertex-base pair this corresponds
naturally to a protected hub and a noisy spoke, a configuration
motivated by hub-and-spoke quantum-network
architectures~\cite{kimble2008quantum}, whereas for symmetric and
base-base pairs it should be understood simply as a one-sided channel
assignment. This convention isolates the channel action from the
geometry of the state and allows the different pair classes to be
compared on the same footing.

The resulting framework lets us revisit the claim of
Ref.~\cite{bhattacharyya2026super}. By separating network geometry
from excitation sector and extending the comparison to arbitrary $N$,
we show that the VB--$W$ reordering arises in the cross-sector
comparison and therefore does not, by itself, establish an intrinsic
fragility of the vertex-base geometry, while the lower ESD threshold
of the base-base link under DP and, within the $(N-1)$-excitation
sector, under AD survives as a distinct same-sector structural effect.

The paper is organized as follows. Section~\ref{sec:prelim} collects
the technical tools: the X-state form of the reduced density matrices,
the Wootters concurrence formula, and the $\ell_{1}$-norm of
coherence. Section~\ref{sec:symmetric} treats the symmetric family and
its network geometry. Section~\ref{sec:channels} specifies the four
noise channels and the one-sided noise convention.
Section~\ref{sec:sym-dyn} gives the decoherence dynamics of the
symmetric family. Section~\ref{sec:asymmetric} introduces the
generalized asymmetric family and develops its weighted
entanglement-network geometry, including a comparison with the
symmetric family. Section~\ref{sec:asym-dyn} presents the decoherence
dynamics of the asymmetric family. Section~\ref{sec:reinterp} revisits
the super-link fragility effect by separating the roles of network
geometry and excitation sector. Section~\ref{sec:disc} synthesises the
comparative picture and Section~\ref{sec:conc} summarises. Supporting
results and figures are collected in Appendix~\ref{app:supporting},
while the full derivations appear in
Appendices~\ref{app:general}--\ref{app:noiseF-VB}.

\section{Preliminaries}
\label{sec:prelim}

\subsection{X-states and the Wootters concurrence}

Every two-qubit reduced density matrix appearing in this work has,
in the computational basis
$\{|00\rangle,|01\rangle,|10\rangle,|11\rangle\}$, the X-state form
\begin{equation}
\label{eq:Xstate}
\rho_{X}=
\begin{pmatrix}
a & 0 & 0 & w\\
0 & b & z & 0\\
0 & z^{*} & c & 0\\
w^{*} & 0 & 0 & d
\end{pmatrix}, \ a+b+c+d=1, \ z,w\in\mathbb{C}.
\end{equation}
The Wootters concurrence~\cite{wootters1998entanglement,hill1997entanglement} of $\rho_{X}$
reduces to (see Appendix~\ref{app:general})
\begin{equation}
\label{eq:Xconc}
\mathcal{C}(\rho_{X})
=2\max\bigl\{0,\ |z|-\sqrt{ad},\ |w|-\sqrt{bc}\bigr\}.
\end{equation}
Every initial state in this work satisfies $w=0$, and every noise
channel preserves this property, so the second branch of
Eq.~\eqref{eq:Xconc} never activates. The working formula throughout
the manuscript is
\begin{equation}
\label{eq:concwork}
\mathcal{C}(\rho_{X})
=2\max\bigl\{0,\ |z|-\sqrt{ad}\bigr\}.
\end{equation}

\subsection{$\ell_{1}$-norm of coherence}

For a pure state $|\psi\rangle=\sum_{i}c_{i}|i\rangle$ in the
computational basis, the $\ell_{1}$-norm of coherence
is~\cite{baumgratz2014quantifying}
\begin{equation}
\label{eq:ell1def}
\Cl(|\psi\rangle)=\Bigl(\sum_{i}|c_{i}|\Bigr)^{2}-1.
\end{equation}

We use $\Cl$ as a state-preparation diagnostic and, in
Appendix~\ref{app:supporting}, as a structural signature that
distinguishes the vertex qubit from the base qubits of the
asymmetric family.

\begin{figure*}[t!]
	\centering
	\includegraphics[width=0.48\textwidth]{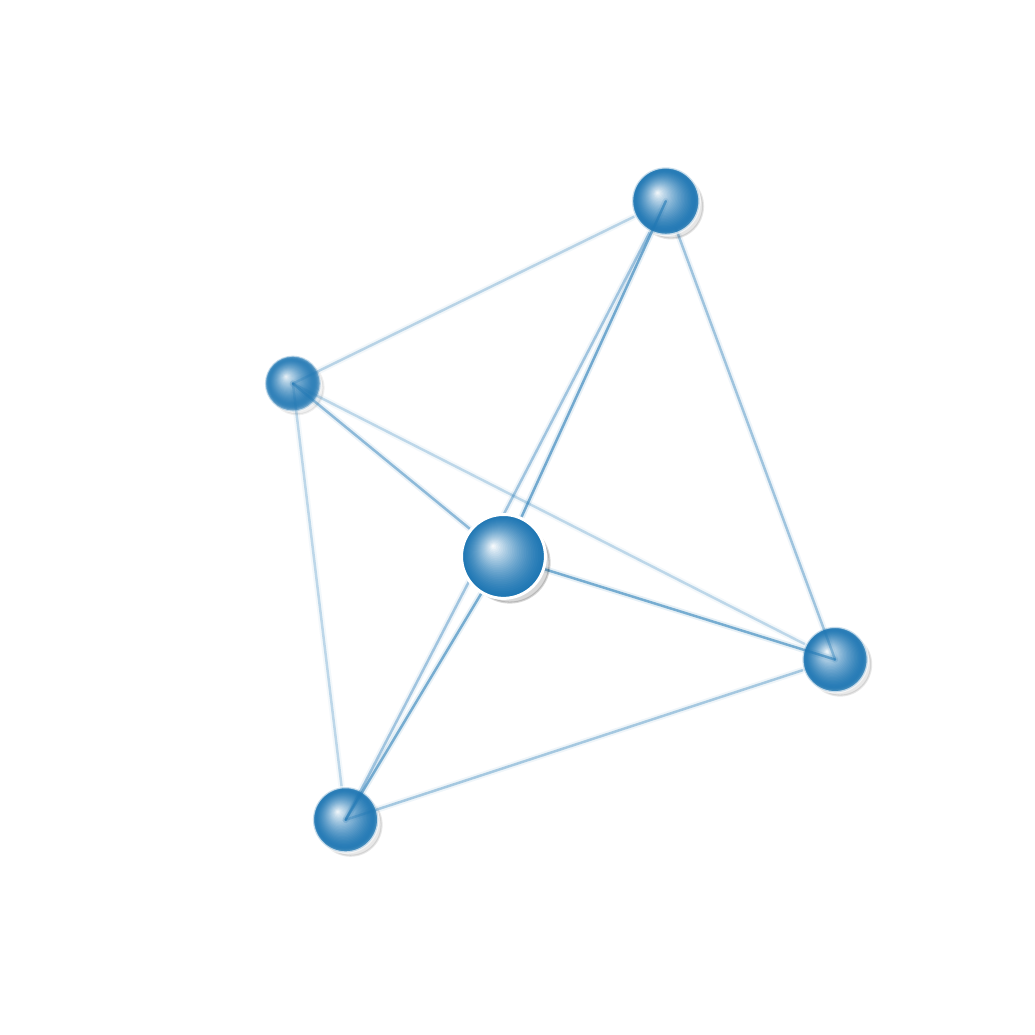} \quad
	\includegraphics[width=0.48\textwidth]{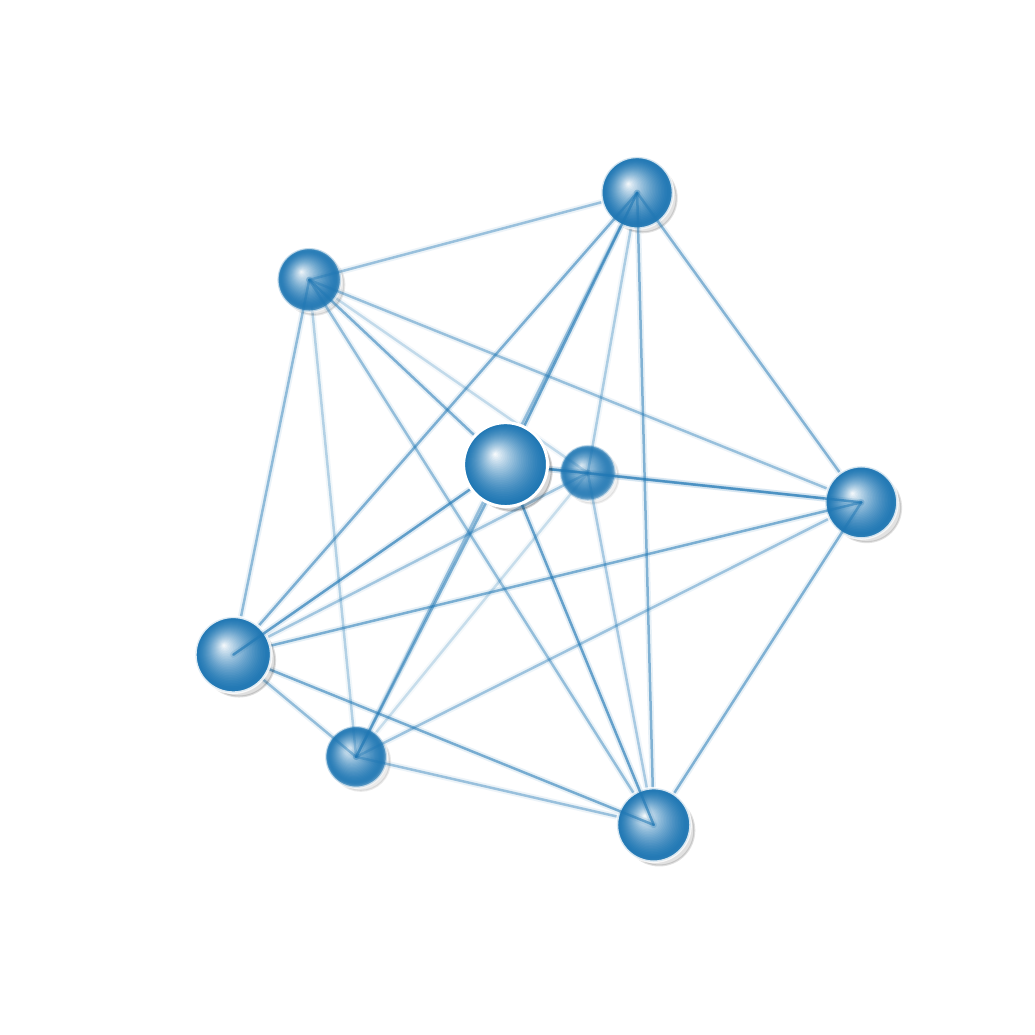}
	\caption{Equilateral entanglement-network geometry of symmetric
		$|W\rangle$ states for $N=5$ (left) and $N=8$ (right). Every pair carries
		concurrence $2/N$, illustrating that all qubits play structurally identical roles with no distinguished node.}
	\label{fig:sym-topology}
\end{figure*}

\section{Symmetric $N$-qubit $|W\rangle$ states and their network
geometry}
\label{sec:symmetric}

\subsection{Definitions}

The symmetric $N$-qubit $|W\rangle$ state is
\begin{equation}
\WN=\frac{1}{\sqrt{N}}\sum_{i=1}^{N}|0\cdots1_{i}\cdots0\rangle,
\label{eq:WN}
\end{equation}
and its global bit-flipped partner is
\begin{equation}
\WbarN=\sigma_{x}^{\otimes N}\WN
=\frac{1}{\sqrt{N}}\sum_{i=1}^{N}|1\cdots0_{i}\cdots1\rangle.
\label{eq:WbarN}
\end{equation}
$\WN$ lies in the single-excitation subspace; $\WbarN$ lies in the
$(N-1)$-excitation subspace. Both are fully permutation-symmetric,
and both have $\Cl=N-1$.

\subsection{Reduced density matrices and initial concurrence}

By permutation symmetry, every bipartite reduction of $\WN$ is
identical. Tracing out qubits $3,\ldots,N$ (Appendix~\ref{app:initial-W})
gives
\begin{eqnarray}
\rho_{12}^{W}=\frac{1}{N}
\begin{pmatrix}
N-2 & 0 & 0 & 0\\
0 & 1 & 1 & 0\\
0 & 1 & 1 & 0\\
0 & 0 & 0 & 0
\end{pmatrix}, \nonumber \\
\rho_{12}^{\bar W}=\frac{1}{N}
\begin{pmatrix}
0 & 0 & 0 & 0\\
0 & 1 & 1 & 0\\
0 & 1 & 1 & 0\\
0 & 0 & 0 & N-2
\end{pmatrix}.
\label{eq:rhoW-rhoWbar}
\end{eqnarray}
Applying Eq.~\eqref{eq:concwork},
\begin{equation}
\mathcal{C}(\rho_{12}^{W})=\mathcal{C}(\rho_{12}^{\bar W})=\frac{2}{N},
\qquad N\ge2.
\label{eq:Cinit-symm}
\end{equation}
The pairwise concurrence decreases inversely with the system size:
larger symmetric $|W\rangle$-class states begin with weaker
per-pair entanglement, spread uniformly over $\binom{N}{2}$ links.

\subsection{Network geometry}

The pairwise-concurrence graph of $\WN$ is the complete graph
$K_{N}$ with all $\binom{N}{2}$ edge weights equal to $2/N$. 
We refer to this as the \emph{equilateral} network. Its defining
feature is that all qubits play structurally identical roles: no
distinguished node exists. Figure~\ref{fig:sym-topology} illustrates
this geometry for representative system sizes.

\section{Local quantum noise channels}
\label{sec:channels}

An open-system evolution is described by the operator-sum
representation
$\rho'=\sum_{i}E_{i}\rho E_{i}^{\dagger}$, with
$\sum_{i}E_{i}^{\dagger}E_{i}=I$.

\paragraph{One-sided noise convention\\}
For every bipartite subsystem, one qubit is protected and one is
noisy:
\begin{equation}
\label{eq:onesided}
E_{k}=I\otimes K_{k}.
\end{equation}
In a VB pair the vertex is protected and the base qubit is noisy;
in a BB pair one base qubit is noisy and the other is protected.
For the symmetric family, the same one-sided assignment is made to
either member of the equivalent pair. This convention matches the
local-noise treatment of our recent three-qubit
work~\cite{bhattacharyya2026super}.

\paragraph{Kraus operators\\}
Phase damping (PD):
\begin{equation}
K_{0}=\begin{pmatrix}1&0\\0&\sqrt{1-p}\end{pmatrix},\ \ 
K_{1}=\begin{pmatrix}0&0\\0&\sqrt{p}\end{pmatrix}, \ \ p\in[0,1].
\label{eq:kraus-PD}
\end{equation}
Amplitude damping (AD):
\begin{equation}
K_{0}=\begin{pmatrix}1&0\\0&\sqrt{1-\gamma}\end{pmatrix},\ \
K_{1}=\begin{pmatrix}0&\sqrt{\gamma}\\0&0\end{pmatrix}, \ \ \gamma\in[0,1].
\label{eq:kraus-AD}
\end{equation}
Depolarization (DP):
\begin{eqnarray}
K_{0}=\sqrt{1-p}\,I,\quad
K_{1}=\sqrt{p/3}\,\sigma_{x},\quad \nonumber \\
K_{2}=\sqrt{p/3}\,\sigma_{y},\quad 
K_{3}=\sqrt{p/3}\,\sigma_{z}.
\label{eq:kraus-DP}
\end{eqnarray}
Here $p\in[0,1]$ is the total Pauli-error probability; the completely
depolarizing point occurs at $p=3/4$. All ESD thresholds obtained
below lie below this point.
Generalized amplitude damping (GAD), with damping probability
$p$ and thermal weight $\alpha\in[0,1]$:
\begin{eqnarray}
K_{0}&=\sqrt{1-\alpha}\begin{pmatrix}\sqrt{1-p}&0\\0&1\end{pmatrix},\  K_{1}&=\sqrt{1-\alpha}\begin{pmatrix}0&0\\\sqrt{p}&0\end{pmatrix},\notag\\
K_{2}&=\sqrt{\alpha}\begin{pmatrix}1&0\\0&\sqrt{1-p}\end{pmatrix}, \  K_{3}&=\sqrt{\alpha}\begin{pmatrix}0&\sqrt{p}\\0&0\end{pmatrix}.
\label{eq:kraus-GAD}
\end{eqnarray}
At $\alpha=1$ this reduces to AD; at $\alpha=0$ it is the
pure-excitation (pumping) limit. This single GAD convention is used
for every state throughout the paper and matches that of our recent
three-qubit work~\cite{bhattacharyya2026super}.

\begin{figure*}[t!]
	\centering
	\includegraphics[width=0.95\textwidth]{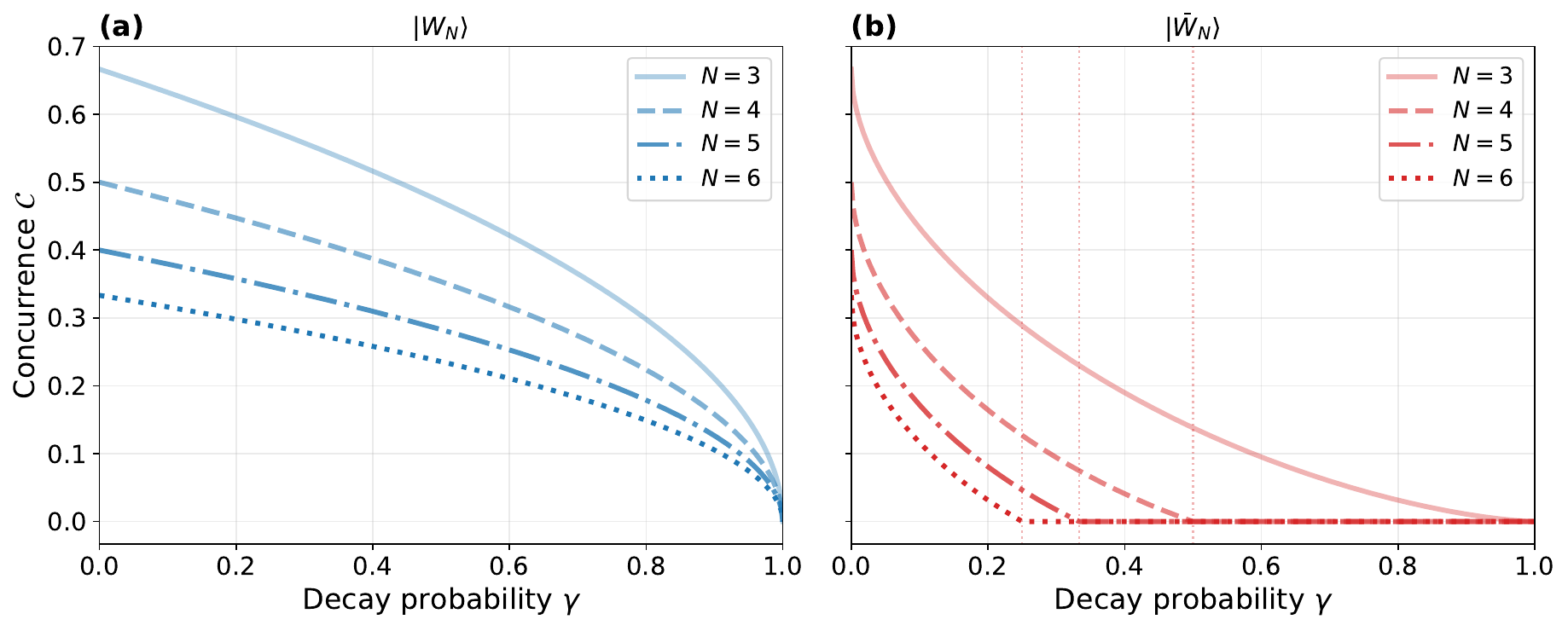}
	\caption{AD concurrence. \emph{Left:} $\WN$ decays without ESD.
		\emph{Right:} $\WbarN$ undergoes ESD at
		$\gamma^{*}=1/(N-2)$ for $N\ge4$; at $N=3$, concurrence vanishes
		only at the boundary $\gamma=1$, so no ESD occurs.}
	\label{fig:ADsym}
\end{figure*}

\section{Decoherence dynamics of symmetric $|W\rangle$-class states}
\label{sec:sym-dyn}

We apply the one-sided noise model of Eq.~\eqref{eq:onesided} to
the reduced density matrices Eq.~\eqref{eq:rhoW-rhoWbar}. The full
derivations are collected in Appendices~\ref{app:noiseC-W}
and~\ref{app:noiseD-W}. We report only the final concurrence
expressions here. Supporting plots for the symmetric-family dynamics
are collected in Appendix~\ref{app:supporting}.

\subsection{Phase damping}

PD suppresses only the off-diagonal element $z$, leaving diagonal
populations unchanged. Hence
\begin{equation}
\mathcal{C}^{W}_{N}(p)=\mathcal{C}^{\bar W}_{N}(p)
=\frac{2\sqrt{1-p}}{N},
\label{eq:CPsym}
\end{equation}
and no ESD occurs for any $N$: PD preserves the
dynamical equivalence between the two bit-flipped partners.

\subsection{Amplitude damping}

AD is directional: spontaneous emission moves population from
$|1\rangle$ to $|0\rangle$. For $\WN$ the $|00\rangle$ population
grows while $\rho_{44}$ stays identically zero throughout, so the
$\sqrt{ad}$ penalty in Eq.~\eqref{eq:concwork} never activates:
\begin{equation}
\mathcal{C}^{W}_{N}(\gamma)=\frac{2\sqrt{1-\gamma}}{N},
\label{eq:CADsym-W}
\end{equation}
with no ESD. For $\WbarN$ the ($N-1$)-excitation structure causes
both corner elements to acquire non-zero population, activating the
penalty:
\begin{equation}
\mathcal{C}^{\bar W}_{N}(\gamma)
=\frac{2\sqrt{1-\gamma}}{N}\max\!\left\{0,\,1-\sqrt{\gamma(N-2)}\right\},
\label{eq:CADsym-Wbar}
\end{equation}
which vanishes at
\begin{equation}
\gamma^{*}=\frac{1}{N-2},\qquad N\ge4.
\label{eq:gammastar}
\end{equation}
At $N=3$, the formal solution $\gamma=1$ lies at the boundary of the
physical range, so no ESD occurs. For $N\ge4$,
larger systems lose bipartite entanglement at progressively smaller $\gamma$, and
$\gamma^{*}\to0$ as $N\to\infty$: the ESD threshold of
$(N-1)$-excitation $|W\rangle$-class states therefore vanishes in
the large-$N$ limit. 
Figure~\ref{fig:ADsym} illustrates the contrasting AD dynamics of the
single- and $(N-1)$-excitation symmetric states.

The contrasting behaviour of $\WN$ and $\WbarN$ under AD --- despite
their identical initial concurrence and coherence --- is the
first appearance in this paper of an \emph{excitation-sector} effect:
the robustness of pairwise entanglement depends not only on the
initial amount of entanglement but also on the excitation content
of the underlying state.

\subsection{Depolarization}

DP is isotropic: both populations and coherences are modified
simultaneously and symmetrically. The concurrence is identical for
$\WN$ and $\WbarN$:
\begin{widetext}
	\begin{equation}
	\mathcal{C}^{W}_{N}(p)=\mathcal{C}^{\bar W}_{N}(p)
	=\frac{2}{N}\max\!\left\{0,\,
	\left|1-\frac{4p}{3}\right|
	-\sqrt{\frac{2p(N-2)}{3}-\frac{4p^{2}(N-3)}{9}}\right\},
	\label{eq:CDPsym}
	\end{equation}
\end{widetext}
	
with common ESD threshold
\begin{equation}
p^{*}=\frac{3}{2(N+1)}.
\label{eq:pDPsym}
\end{equation}
The threshold decreases monotonically with $N$ and vanishes as
$N\to\infty$. Thus, for any fixed $p>0$ within the depolarizing
regime $p<3/4$, sufficiently large systems undergo pairwise ESD.

\subsection{Generalized amplitude damping}

The GAD concurrences of $\WN$ and $\WbarN$ are
\begin{widetext}
\begin{align}
\mathcal{C}^{W}_{N}(p,\alpha)
&=\frac{2}{N}\max\!\left\{0,\,\sqrt{1-p}
-\sqrt{(1-\alpha)p\bigl[N-2+p\bigl(\alpha(N-1)-(N-2)\bigr)\bigr]}\,\right\},
\label{eq:CGADsym-W}\\[2pt]
\mathcal{C}^{\bar W}_{N}(p,\alpha)
&=\frac{2}{N}\max\!\left\{0,\,\sqrt{1-p}
-\sqrt{\alpha p\bigl[N-2+p\bigl(1-\alpha(N-1)\bigr)\bigr]}\,\right\}.
\label{eq:CGADsym-Wbar}
\end{align}
\end{widetext}
The substitution $\alpha\to1-\alpha$ maps
Eq.~\eqref{eq:CGADsym-W} onto Eq.~\eqref{eq:CGADsym-Wbar} exactly:
\begin{equation}
\mathcal{C}^{W}_{N}(p,\alpha)=\mathcal{C}^{\bar W}_{N}(p,1-\alpha),
\label{eq:bitflip-cov-sym}
\end{equation}
which expresses the exact bit-flip covariance of the concurrence
dynamics under $\alpha\leftrightarrow1-\alpha$. Its physical content is
that the global bit-flip operation $\sigma_{x}^{\otimes N}$
connecting $\WN$ and $\WbarN$ is mirrored, at the level of the GAD
Kraus operators, by the exchange of relaxation and excitation
channels.

Two limits follow immediately. At $\alpha=1$ (pure AD), $\WN$ has
no ESD, while $\WbarN$ undergoes ESD at
$p^{*}=1/(N-2)$ for $N\ge4$; at $N=3$, concurrence vanishes only at
the boundary $p=1$. At $\alpha=0$ (pure excitation), the roles of
$\WN$ and $\WbarN$ interchange.
The symmetry point between the two regimes occurs at
\begin{equation}
	\alpha^{*}=\tfrac{1}{2},
	\label{eq:alphastar}
\end{equation}
at which the two states have identical concurrence dynamics for
every $N$ and every $p$. GAD therefore continuously interpolates
between relaxation- and excitation-dominated regimes, with the two
bit-flipped partners related exactly by $\alpha\leftrightarrow1-\alpha$.

\subsection{ESD threshold scaling with system size}

\begin{figure}[b!]
\centering
\includegraphics[width=0.5\textwidth]{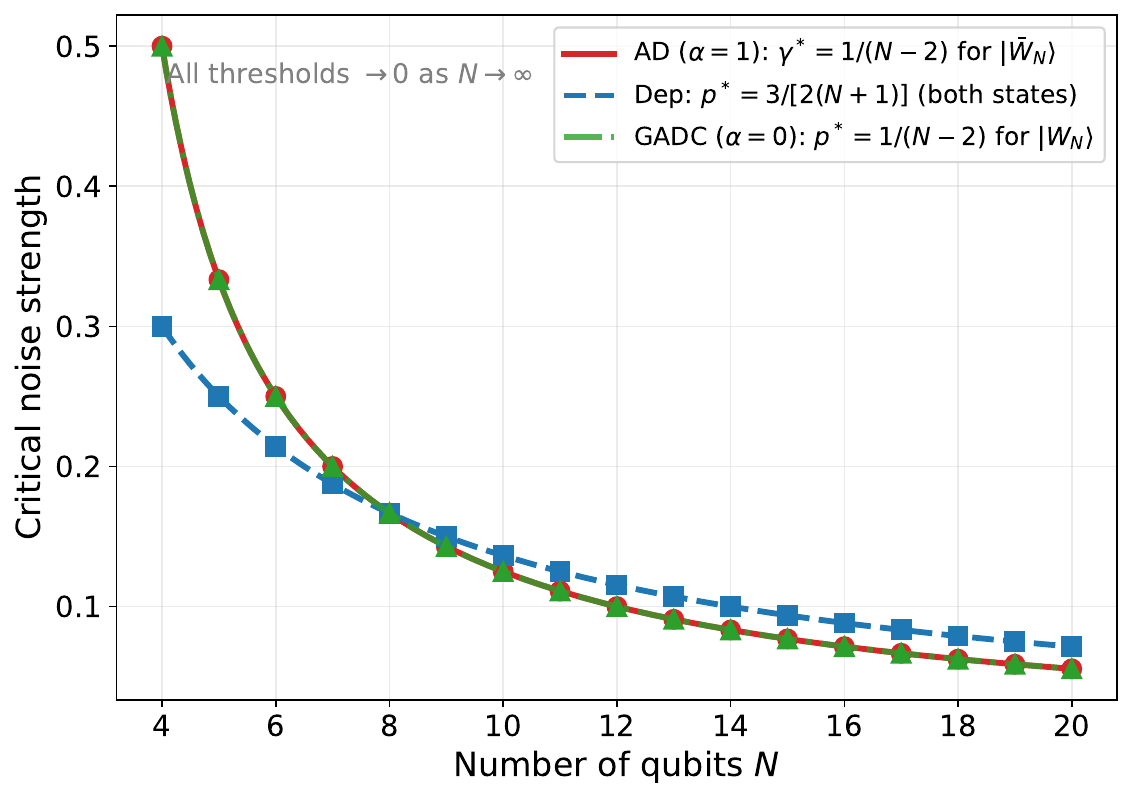}
\caption{ESD thresholds for the symmetric family under
	AD and DP. The AD threshold is $\gamma^{*}=1/(N-2)$ for $N\ge4$,
	while the DP threshold is $p^{*}=3/[2(N+1)]$; both vanish as
	$N\to\infty$.}
\label{fig:ESDsymN}
\end{figure}

Figure~\ref{fig:ESDsymN} compares the system-size dependence of the
symmetric-family ESD thresholds under AD and DP.

\section{Generalized asymmetric $N$-qubit $W$-class states}
\label{sec:asymmetric}

\subsection{Definitions}

The generalized asymmetric $N$-qubit $|W\rangle$-class state is
\begin{equation}
\WL=\frac{1}{\sqrt{2}}\bigl|1\,\underbrace{0\,0\cdots0}_{N-1}\bigr\rangle
+\frac{1}{\sqrt{2(N-1)}}\sum_{k=2}^{N}|0\cdots1_{k}\cdots0\rangle,
\label{eq:WL}
\end{equation}
which lies in the single-excitation subspace, and its bit-flipped
partner
\begin{align}
\WLb=&\sigma_{x}^{\otimes N}\WL \nonumber \\ 
=&\frac{1}{\sqrt{2}}\bigl|0\,\underbrace{1\,1\cdots1}_{N-1}\bigr\rangle \nonumber \\
+&\frac{1}{\sqrt{2(N-1)}}\sum_{k=2}^{N}|1\cdots0_{k}\cdots1\rangle,
\label{eq:WLb}
\end{align}
which lies in the $(N-1)$-excitation subspace. Qubit~1 is the
\emph{vertex}, qubits $2,\ldots,N$ are the $(N-1)$ equivalent
\emph{base} qubits.

\paragraph{Relation to the Lohmayer state\\}
\label{obs:lohmayer}
At $N=3$, $\WL$ reduces to
$\tfrac{1}{\sqrt{2}}|100\rangle+\tfrac{1}{2}|010\rangle
+\tfrac{1}{2}|001\rangle$, which is the single-excitation global
bit-flipped partner of the Lohmayer state~\cite{lohmayer2006entangled}.
Conversely, $\WLb$ at $N=3$ reduces to
$\tfrac{1}{\sqrt{2}}|011\rangle+\tfrac{1}{2}|101\rangle
+\tfrac{1}{2}|110\rangle$, which is the two-excitation Lohmayer
state used in~\cite{bhattacharyya2026super}. This distinction is
physically consequential because AD is directional: same-sector and
cross-sector comparisons are inequivalent at $N=3$.

\subsection{Reduced density matrices and initial concurrence}

The vertex qubit is inequivalent to any base qubit, so two
distinct bipartite reductions arise: $\rho_{VB}\equiv\rho_{12}$
(vertex-base) and $\rho_{BB}\equiv\rho_{23}$ (base-base). Explicit
computation (Appendix~\ref{app:initial-VB}) gives
\begin{eqnarray}
\rho_{12}^{W^{L}}=
\begin{pmatrix}
\frac{N-2}{2(N-1)} & 0 & 0 & 0\\
0 & \frac{1}{2(N-1)} & \frac{1}{2\sqrt{N-1}} & 0\\
0 & \frac{1}{2\sqrt{N-1}} & \frac{1}{2} & 0\\
0 & 0 & 0 & 0
\end{pmatrix},\nonumber \\
\rho_{23}^{W^{L}}=
\begin{pmatrix}
\frac{N-2}{N-1} & 0 & 0 & 0\\
0 & \frac{1}{2(N-1)} & \frac{1}{2(N-1)} & 0\\
0 & \frac{1}{2(N-1)} & \frac{1}{2(N-1)} & 0\\
0 & 0 & 0 & 0
\end{pmatrix}.
\label{eq:rhoVB-BB-WL}
\end{eqnarray}
By Eq.~\eqref{eq:concwork},
\begin{eqnarray}
\CVB(W^{L})=\frac{1}{\sqrt{N-1}}=\CVB(\overline{W^{L}}),\nonumber \\
\CBB(W^{L})=\frac{1}{N-1}=\CBB(\overline{W^{L}}),
\label{eq:Cinit-asym}
\end{eqnarray}
where the second equalities follow from the global bit-flip
symmetry that swaps $a\leftrightarrow d$ (and leaves $|z|$
invariant) in Eq.~\eqref{eq:concwork}. Combining with
Eq.~\eqref{eq:Cinit-symm} gives the following exact structural
result.

\paragraph{Concurrence hierarchy\\}
\label{res:hierarchy}
For all $N\ge3$,
\begin{equation}
\CVB=\frac{1}{\sqrt{N-1}}\ >\ \CW=\frac{2}{N}\ >\
\CBB=\frac{1}{N-1}.
\label{eq:hierarchy}
\end{equation}

\begin{figure}[t!]
\centering
\includegraphics[width=0.5\textwidth]{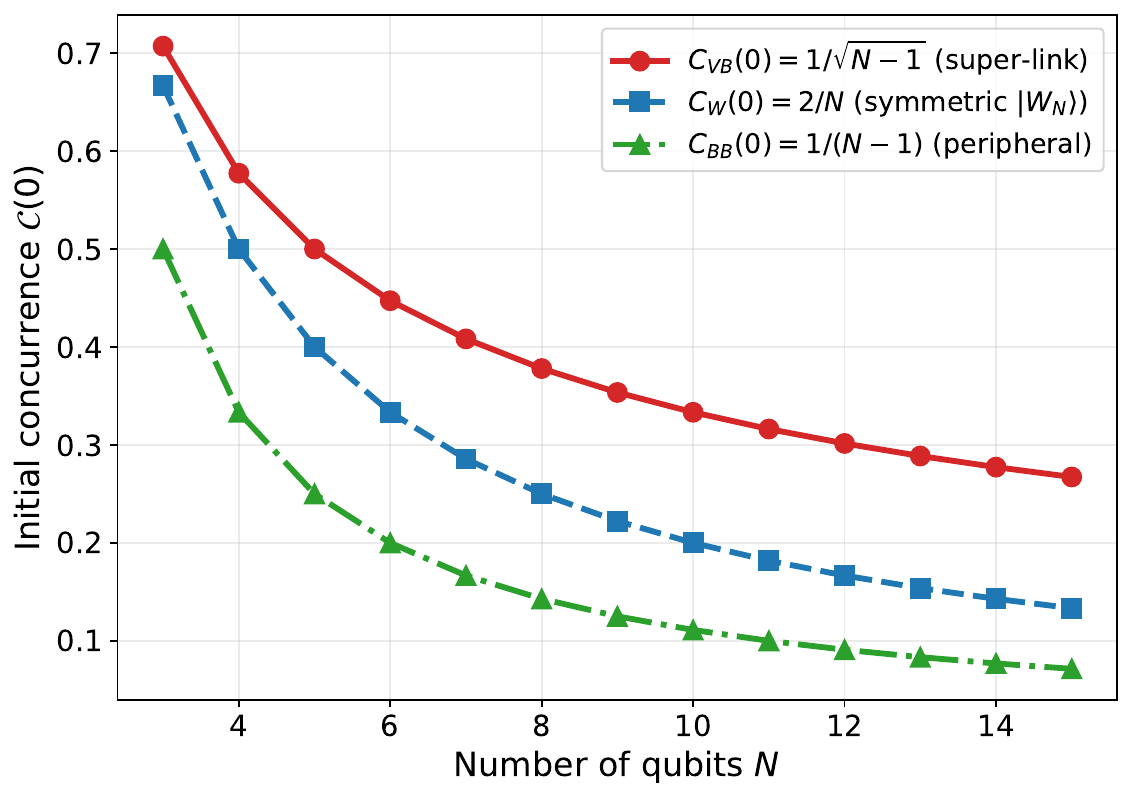}
\caption{Initial pairwise concurrence versus $N$ for the VB and BB
	pair classes and the symmetric reference. The hierarchy
	$\CVB>\CW>\CBB$ holds for all $N\ge3$.}
\label{fig:hierarchy}
\end{figure}

The three concurrences scale differently: $\CVB\sim N^{-1/2}$,
$\CW\sim N^{-1}$, $\CBB\sim (N-1)^{-1}$. The vertex-base pair class has the largest pairwise concurrence in
the asymmetric network, while the base-base pair class has the
smallest; the symmetric reference lies between them. Figure~\ref{fig:hierarchy} illustrates this initial concurrence hierarchy and its scaling with system size. 

\subsection{Weighted entanglement-network geometry}
\label{sec:topology}

\begin{figure*}[t!]
	\centering
	\includegraphics[width=0.48\textwidth]{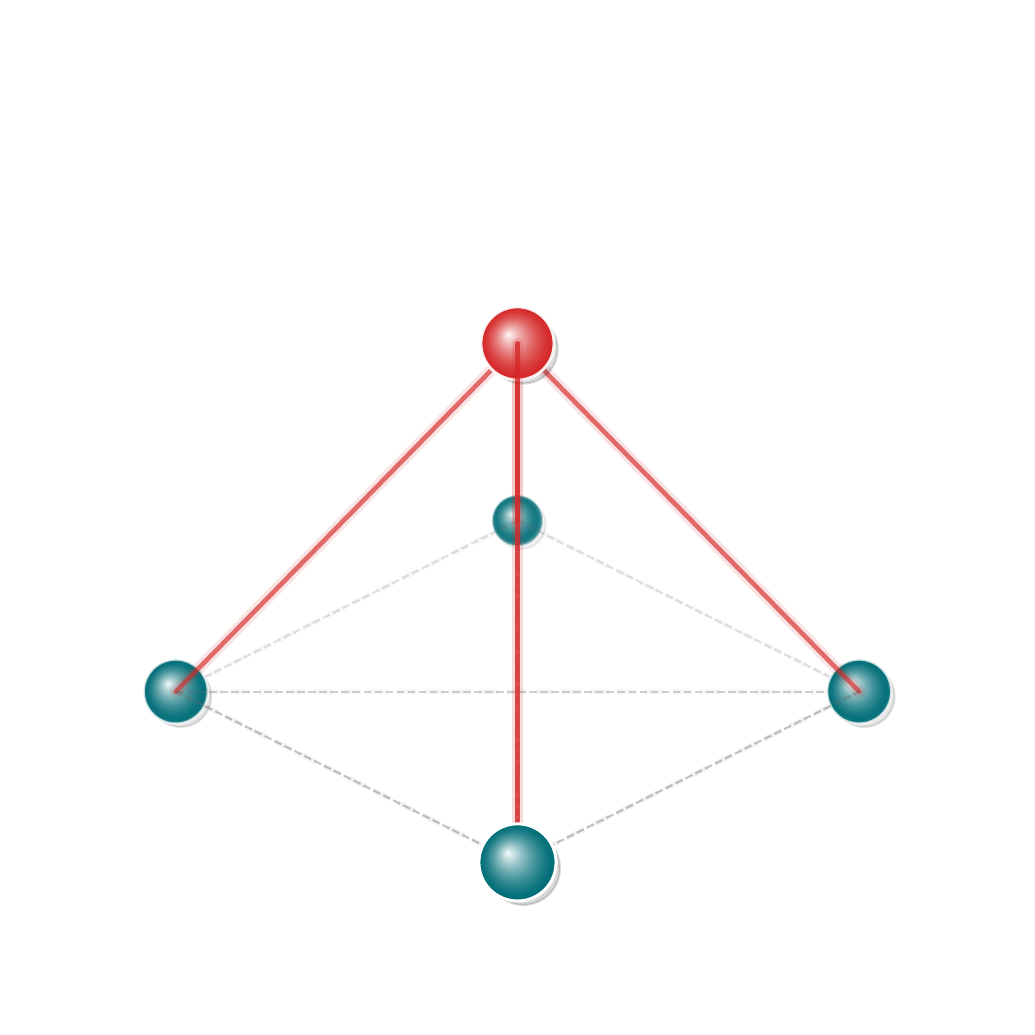}\hfill
	\includegraphics[width=0.48\textwidth]{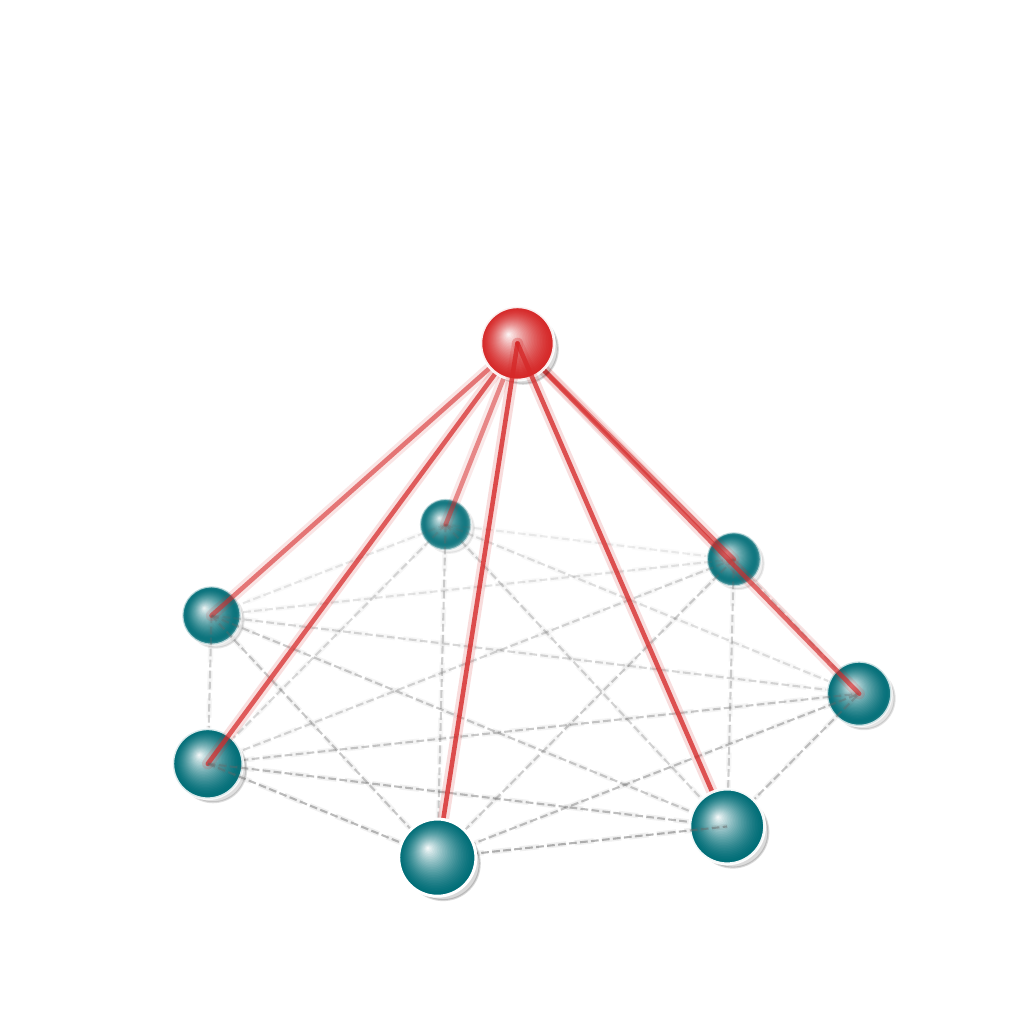}
	\caption{Entanglement-network geometry of the generalized
		asymmetric $\WL$ state for $N=5$ (left) and $N=8$ (right). The distinguished vertex
		qubit is connected to each base qubit by a higher-concurrence VB link
		($\CVB=1/\sqrt{N-1}$, solid), while the base qubits are mutually
		connected by lower-concurrence BB links ($\CBB=1/(N-1)$, dashed). The overall
		architecture is a star-plus-clique on $N$ nodes.}
	\label{fig:asym-topology}
\end{figure*}

\paragraph{The star-plus-clique architecture}

The pairwise-concurrence network of the generalized asymmetric
$N$-qubit $|W\rangle$-class states has a \emph{star-plus-clique}
architecture. The underlying graph is complete, but its edges fall
into two inequivalent concurrence-weight classes, corresponding to
the vertex-base and base-base pair classes.

First, the \emph{star} component consists of a distinguished central hub, referred to as the vertex qubit, which is connected to each of the $(N-1)$ equivalent base qubits. These $(N-1)$ vertex-base edges carry the largest pairwise concurrence in the network, $\mathcal{C}_{VB} = 1/\sqrt{N-1}$.

Second, the \emph{clique} component is formed by the complete interconnection of these peripheral base qubits. This results in a fully connected subgraph comprising $\binom{N-1}{2}$ base-base edges, each carrying pairwise concurrence $\mathcal{C}_{BB} = 1/(N-1)$. 

Taken together, the network functions as a centralized hub-and-spoke system that is augmented by a fully connected peripheral mesh.

For $N=3$ the base clique reduces to a single BB edge and the
network is an isosceles triangle, recovering the three-qubit
Lohmayer geometry of~\cite{bhattacharyya2026super}. For $N=4$ the base
clique becomes a triangle, so the total network is a $K_{4}$ with
three edges of one weight and three of another. For $N\ge5$ the
base clique becomes a $K_{N-1}$ subgraph, and the total network is
a hub-and-spoke architecture augmented by a complete set of
base-base concurrence edges. In every case, the vertex remains a distinguished
node. For the VB pair class, this makes the protected-vertex/noisy-base
assignment of Section~\ref{sec:channels} a natural realization of
the one-sided local-noise convention. Figure~\ref{fig:asym-topology}
illustrates this star-plus-clique entanglement-network geometry for
representative system sizes.

\paragraph{Contrast with the equilateral symmetric network\\}
The symmetric state $\WN$ has $K_{N}$ with all $\binom{N}{2}$ edge
weights equal to $2/N$: no distinguished node, no distinguished
edge, a single per-pair concurrence. The asymmetric state has the
same underlying complete graph but two concurrence weights; one node
(the vertex) sits on all $(N-1)$ higher-concurrence edges and none
of the lower-concurrence edges; each base node sits on one
higher-concurrence edge and $(N-2)$ lower-concurrence edges. 
This is the network-theoretic content of ``asymmetry'' in this
family. Two structural consequences follow immediately. First,
tracing out the vertex or a base qubit does not yield the same
reduced state, unlike in the symmetric family, so the distinguished
role of the vertex is directly encoded in the reduced-state
structure. Second, the response to local noise can depend on which pair class, VB or BB, is considered, as demonstrated in Section~\ref{sec:asym-dyn}.

\section{Decoherence dynamics of the asymmetric family}
\label{sec:asym-dyn}

All derivations are collected in
Appendices~\ref{app:noiseC-VB}--\ref{app:noiseF-VB}. We report the
concurrence expressions and ESD thresholds. Supporting plots for the
asymmetric-family dynamics are collected in
Appendix~\ref{app:supporting}.

\subsection{Phase damping}

PD affects only the coherence element $z$. Both pair classes
of both states retain their X-state form and their $\sqrt{1-p}$
dependence:
\begin{eqnarray}
\mathcal{C}^{W^{L}}_{VB}(p)=\mathcal{C}^{\overline{W^{L}}}_{VB}(p)
=\frac{\sqrt{1-p}}{\sqrt{N-1}},\nonumber \\
\mathcal{C}^{W^{L}}_{BB}(p)=\mathcal{C}^{\overline{W^{L}}}_{BB}(p)
=\frac{\sqrt{1-p}}{N-1}.
\label{eq:CPasym}
\end{eqnarray}
Since $\rho_{44}$ vanishes for $\WL$ and $\rho_{11}$ vanishes for
$\WLb$, the $\sqrt{ad}$ penalty in Eq.~\eqref{eq:concwork} is
identically zero throughout, and no ESD occurs for any
$N$. The initial hierarchy is therefore preserved for all $p<1$.

\subsection{Amplitude damping}

For $\WL$ (single excitation) the $\rho_{44}$ element of both VB
and BB reductions remains identically zero throughout the
evolution, so no ESD occurs:
\begin{equation}
\mathcal{C}^{W^{L}}_{VB}(\gamma)=\frac{\sqrt{1-\gamma}}{\sqrt{N-1}},\qquad
\mathcal{C}^{W^{L}}_{BB}(\gamma)=\frac{\sqrt{1-\gamma}}{N-1}.
\label{eq:CADasym-single}
\end{equation}

\begin{figure*}[t!]
	\centering
	\includegraphics[width=0.95\textwidth]{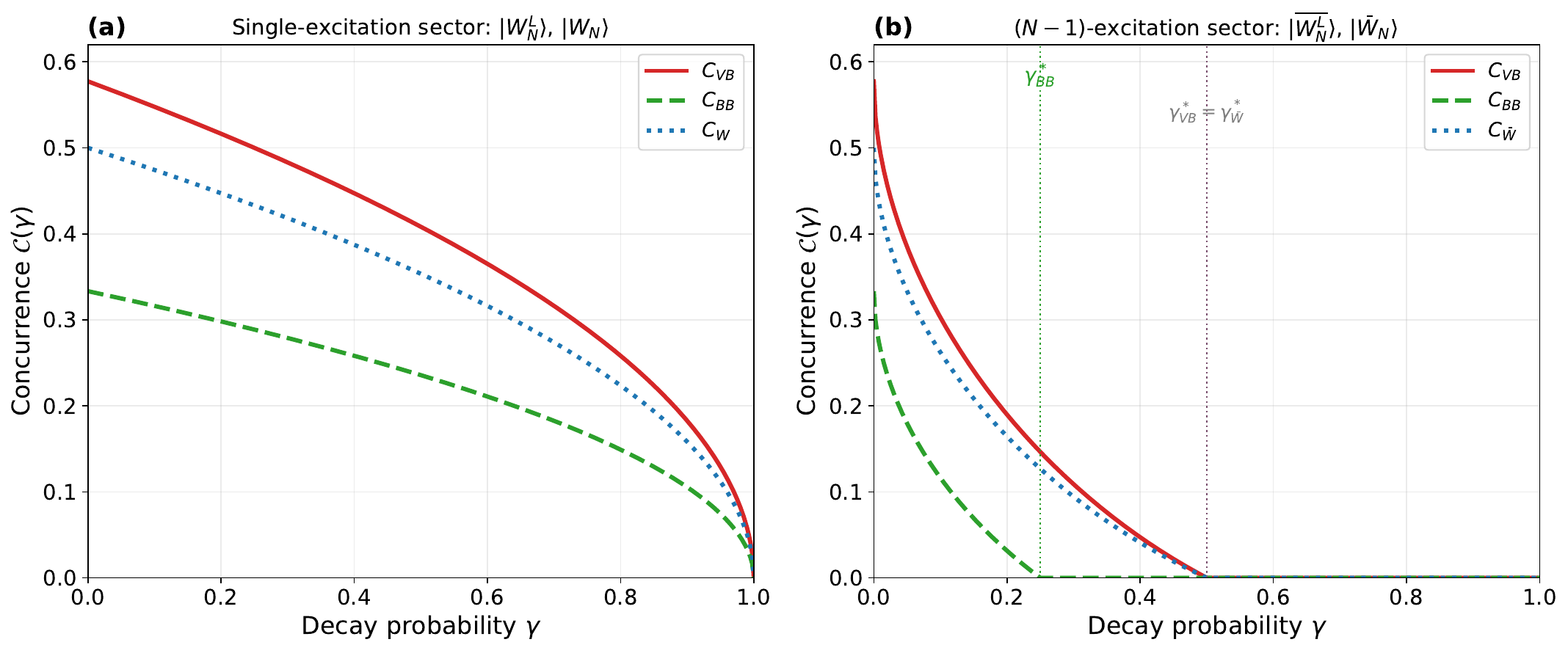}
	\caption{AD concurrence ($N=4$), split by excitation sector.
		\emph{Left:} single-excitation sector $\{\WL,\WN\}$; no ESD occurs
		for the VB or BB pair classes or for the symmetric reference; the
		hierarchy $\CVB>\CW>\CBB$ is preserved for $0\le\gamma<1$, with a
		constant ratio between curves. \emph{Right:} $(N-1)$-excitation
		sector $\{\WLb,\WbarN\}$; the VB and BB pair classes and the
		symmetric reference all exhibit ESD; the constant ratio between $\CVB$ and $\CW$ persists;
		BB undergoes ESD at $\gamma^{*}_{BB}=1/[2(N-2)]$, while VB and
		$\bar W$ undergo ESD together at $\gamma^{*}_{VB}=\gamma^{*}_{\bar W}=1/(N-2)$.}
	\label{fig:ADasym}
\end{figure*}

For $\WLb$ ($(N-1)$ excitations), AD populates both corner
elements of both reductions:

\begin{align}
\mathcal{C}^{\overline{W^{L}}}_{VB}(\gamma)
&=\frac{\sqrt{1-\gamma}}{\sqrt{N-1}}
\max\!\left\{0,\,1-\sqrt{\gamma(N-2)}\right\},
\label{eq:CADasym-VB-Wb}\\
\mathcal{C}^{\overline{W^{L}}}_{BB}(\gamma)
&=\frac{\sqrt{1-\gamma}}{N-1}
\max\!\left\{0,\,1-\sqrt{2\gamma(N-2)}\right\},
\label{eq:CADasym-BB-Wb}
\end{align}
with ESD thresholds

\begin{equation}
\gamma^{*}_{VB}=\frac{1}{N-2},\qquad
\gamma^{*}_{BB}=\frac{1}{2(N-2)}.
\label{eq:gammastar-asym}
\end{equation}

At $N=3$, the formal VB solution $\gamma_{VB}=1$ lies at the boundary
of the physical range, so only the BB pair undergoes ESD; for
$N\ge4$, both $\gamma^{*}_{VB}$ and $\gamma^{*}_{BB}$ are interior
ESD thresholds. The relation
$\gamma^{*}_{BB}=\tfrac{1}{2}\gamma^{*}_{VB}$ holds algebraically for
every $N\ge3$, with the $N=3$ VB value corresponding only to
boundary-point disentanglement. Figure~\ref{fig:ADasym} illustrates
these contrasting AD dynamics for the two excitation sectors at $N=4$.

\paragraph{Constant vertex-base advantage within a sector\\}

A structural observation deserves emphasis. Within either
excitation sector, the vertex-base link and the symmetric reference
concurrence share the same noise prefactor, and their ratio is a
number that depends only on $N$:

\begin{equation}
\frac{\mathcal{C}^{W^{L}}_{VB}(\gamma)}{\mathcal{C}^{W}_{N}(\gamma)}
=\frac{\mathcal{C}^{\overline{W^{L}}}_{VB}(\gamma)}
{\mathcal{C}^{\bar W}_{N}(\gamma)}
=\frac{N}{2\sqrt{N-1}},
\label{eq:const-ratio}
\end{equation}

for all values at which both concurrences are nonzero.

The vertex-base link therefore retains its proportional advantage
over the symmetric reference within the same excitation sector
throughout the interval in which both concurrences are nonzero under
AD. Any apparent crossing between $\CVB$ and $\CW$ under AD must
consequently involve a cross-sector comparison. We return to this
point in Section~\ref{sec:reinterp}.\\

\subsection{Depolarization}

Applying the one-sided DP channel to both VB and BB reductions of
$\WL$ and $\WLb$ gives, in each case, identical concurrences for
the two bit-flipped partners:

\begin{widetext}
\begin{align}
\mathcal{C}^{W^{L}}_{VB}(p)=\mathcal{C}^{\overline{W^{L}}}_{VB}(p)
&=\frac{1}{\sqrt{N-1}}\max\!\left\{0,\,
\left|1-\frac{4p}{3}\right|
-\sqrt{\frac{2p\bigl[3(N-2)-2p(N-3)\bigr]}{9}}\right\},
\label{eq:CDPasym-VB}\\
\mathcal{C}^{W^{L}}_{BB}(p)=\mathcal{C}^{\overline{W^{L}}}_{BB}(p)
&=\frac{1}{N-1}\max\!\left\{0,\,
\left|1-\frac{4p}{3}\right|
-\frac{2}{3}\sqrt{p\bigl[3(N-2)+p(5-2N)\bigr]}\right\},
\label{eq:CDPasym-BB}
\end{align}
\end{widetext}
\ \\

with ESD thresholds
\begin{equation}
p^{*}_{VB}=\frac{3}{2(N+1)},\qquad
p^{*}_{BB}=\frac{3}{2(2N-1)}.
\label{eq:pDPasym}
\end{equation}

\begin{figure}[t!]
\centering
\includegraphics[width=0.5\textwidth]{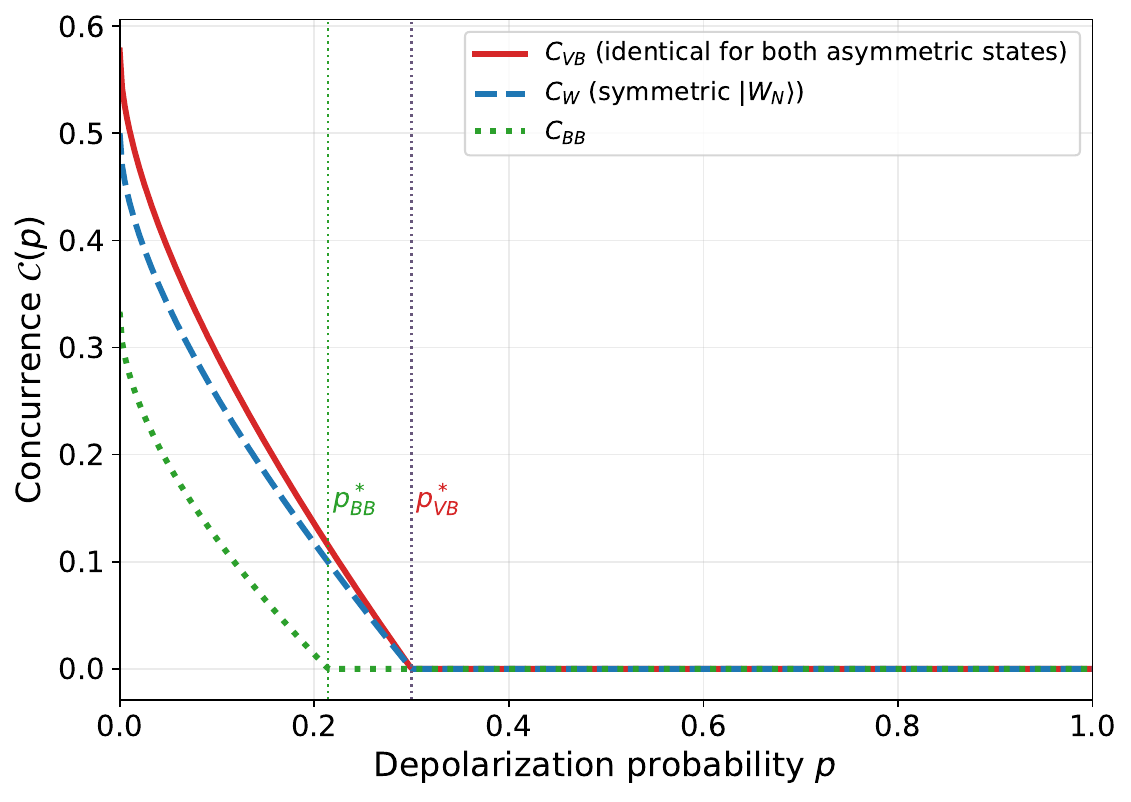}
\caption{DP concurrence ($N=4$). The VB and BB pair classes and the
	symmetric reference all exhibit ESD; the BB pair undergoes ESD first
	because $p^{*}_{BB}=3/[2(2N-1)]<p^{*}_{VB}=3/[2(N+1)]$. Here $p$ is
	the total Pauli-error probability, with complete depolarization at
	$p=3/4$.}
\label{fig:DPasym}
\end{figure}

Since $p^{*}_{BB}<p^{*}_{VB}$ for all $N\ge3$, the BB pair undergoes
ESD before the VB pair. Within each pair class, DP gives identical
concurrence dynamics for the two bit-flipped partners, consistent
with the symmetry already seen in Eq.~\eqref{eq:pDPsym}. Figure~\ref{fig:DPasym} illustrates these depolarization dynamics for
the asymmetric pair classes at $N=4$.

\subsection{Generalized amplitude damping}
\begin{widetext}
For $\WL$:
\begin{align}
\mathcal{C}^{W^{L}}_{VB}(p,\alpha)
&=\frac{1}{\sqrt{N-1}}\max\!\left\{0,\,\sqrt{1-p}
-\sqrt{(1-\alpha)p\bigl[N-2+p\bigl(\alpha(N-1)-(N-2)\bigr)\bigr]}\,\right\},
\label{eq:CGVB-WL}\\
\mathcal{C}^{W^{L}}_{BB}(p,\alpha)
&=\frac{1}{N-1}\max\!\left\{0,\,\sqrt{1-p}
-\sqrt{(1-\alpha)p\bigl[2(N-2)(1-p)+\alpha p(2N-3)\bigr]}\,\right\}.
\label{eq:CGBB-WL}
\end{align}
For $\WLb$:
\begin{align}
\mathcal{C}^{\overline{W^{L}}}_{VB}(p,\alpha)
&=\frac{1}{\sqrt{N-1}}\max\!\left\{0,\,\sqrt{1-p}
-\sqrt{\alpha p\bigl[N-2+p\bigl(1-\alpha(N-1)\bigr)\bigr]}\,\right\},
\label{eq:CGVB-WLbar}\\
\mathcal{C}^{\overline{W^{L}}}_{BB}(p,\alpha)
&=\frac{1}{N-1}\max\!\left\{0,\,\sqrt{1-p}
-\sqrt{\alpha p\bigl[2(N-2)+p\bigl(1-\alpha(2N-3)\bigr)\bigr]}\,\right\}.
\label{eq:CGBB-WLbar}
\end{align}
\end{widetext}
The substitution $\alpha\to1-\alpha$ maps
Eqs.~\eqref{eq:CGVB-WL}--\eqref{eq:CGBB-WL} onto
Eqs.~\eqref{eq:CGVB-WLbar}--\eqref{eq:CGBB-WLbar} exactly:
\begin{eqnarray}
\mathcal{C}^{W^{L}}_{VB}(p,\alpha)
=\mathcal{C}^{\overline{W^{L}}}_{VB}(p,1-\alpha),\nonumber \\
\mathcal{C}^{W^{L}}_{BB}(p,\alpha)
=\mathcal{C}^{\overline{W^{L}}}_{BB}(p,1-\alpha).
\label{eq:bitflip-cov-asym}
\end{eqnarray}

Comparison with Eqs.~\eqref{eq:CGADsym-W}--\eqref{eq:CGADsym-Wbar}
also shows that the VB concurrence and the corresponding same-sector
symmetric concurrence have identical GAD dependence:
\begin{equation}
	\frac{\mathcal{C}^{W^{L}}_{VB}(p,\alpha)}
	{\mathcal{C}^{W}_{N}(p,\alpha)}
	=
	\frac{\mathcal{C}^{\overline{W^{L}}}_{VB}(p,\alpha)}
	{\mathcal{C}^{\bar W}_{N}(p,\alpha)}
	=
	\frac{N}{2\sqrt{N-1}},
	\label{eq:const-ratio-GAD}
\end{equation}

for all values at which both concurrences are nonzero.

At $\alpha=1$ (pure AD), the pairs of $\WL$ have no ESD. For $\WLb$,
the BB pair undergoes ESD at $p^{*}_{BB}=1/[2(N-2)]$ for all
$N\ge3$, while the VB pair undergoes ESD at
$p^{*}_{VB}=1/(N-2)$ for $N\ge4$; at $N=3$, the VB concurrence
vanishes only at the boundary $p=1$. At $\alpha=0$ the roles of
$\WL$ and $\WLb$ interchange. At $\alpha^{*}=\tfrac{1}{2}$ their
corresponding VB and BB concurrence dynamics are identical for every
$N\ge3$ and every $p\in[0,1]$.

Comparing Eqs.~\eqref{eq:CGVB-WL}--\eqref{eq:CGBB-WLbar} across VB
and BB shows that, for every $N\ge3$, the BB square-root penalty is
never smaller than the corresponding VB penalty. The BB radicand
minus the VB radicand is
$(1-\alpha)p(N-2)[1-p(1-\alpha)]$ for $\WL$ and
$\alpha p(N-2)(1-\alpha p)$ for $\WLb$, both nonnegative on
$(p,\alpha)\in[0,1]^2$. Hence the VB entangled region contains the BB
entangled region and is strictly larger overall.

\subsection{ESD threshold scaling with system size}

\begin{figure}[t!]
\centering
\includegraphics[width=0.5\textwidth]{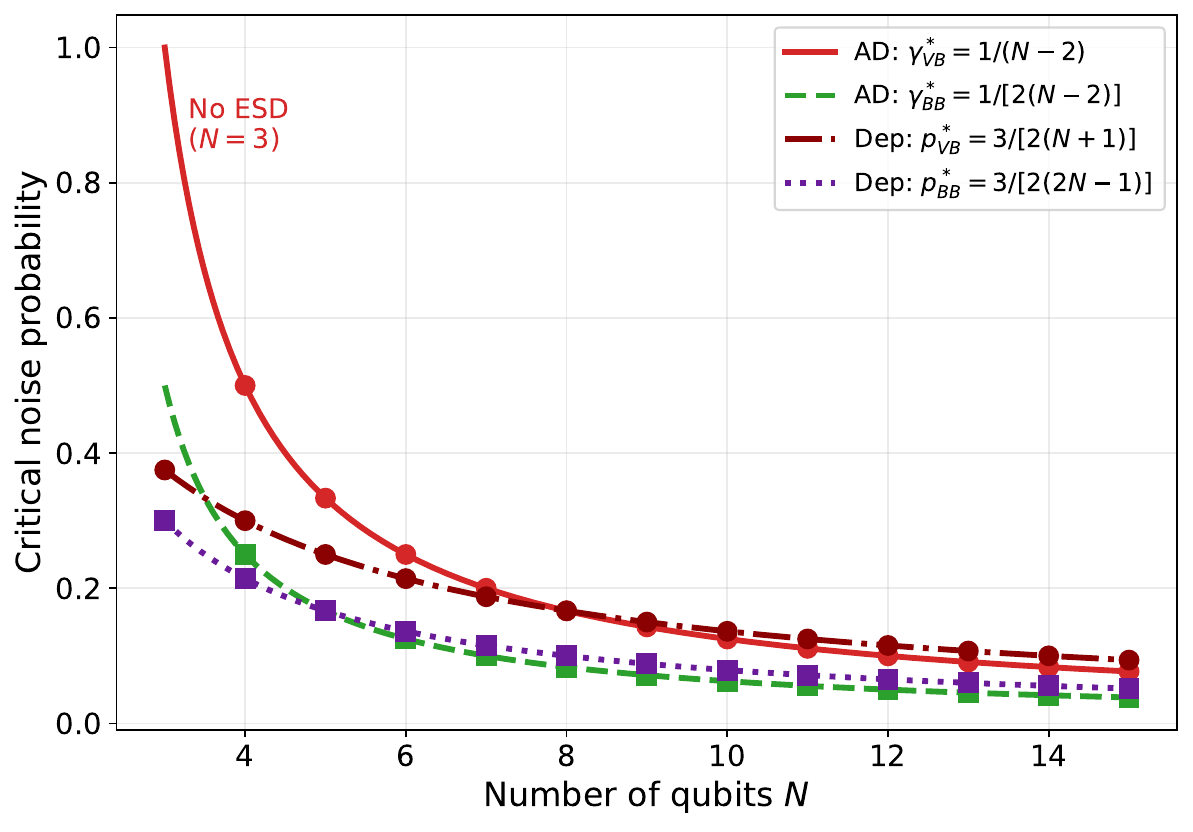}
\caption{ESD threshold scaling with $N$ for the asymmetric family.
	Under DP, the VB threshold exceeds the BB threshold for all $N\ge3$;
	under AD, the same ordering applies to the $(N-1)$-excitation state.}
\label{fig:ESDasymN}
\end{figure}

Figure~\ref{fig:ESDasymN} compares the system-size dependence of the
VB and BB ESD thresholds under AD and DP.

\section{Super-link fragility revisited: separating geometry and excitation sector}
\label{sec:reinterp}

The three-qubit analysis of~\cite{bhattacharyya2026super} reported that,
under amplitude damping, the initial hierarchy $\CVB>\CW>\CBB$ is
overtaken: the vertex-base link loses its advantage over the
symmetric reference and the base-base link undergoes ESD at
$\gamma=1/2$. This was termed the \emph{Super-Link Fragility
Effect}. The $N$-qubit extension developed here makes it possible
to distinguish the mechanisms contributing to this reordering and
to determine which features persist at arbitrary $N$.

\subsection{Excitation-sector origin of the VB--W reordering}

The key structural identity is Eq.~\eqref{eq:const-ratio}: within a
fixed excitation sector, the ratio between the vertex-base
concurrence and the symmetric concurrence is
\begin{equation}
\frac{\mathcal{C}_{VB}(\gamma)}{\mathcal{C}_{W}(\gamma)}
=\frac{N}{2\sqrt{N-1}},
\end{equation}
a constant depending only on $N$, never on the noise strength
$\gamma$. This holds for both the single-excitation pair
$\{\WL,\WN\}$ and the $(N-1)$-excitation pair $\{\WLb,\WbarN\}$
separately. Consequently, \emph{no crossing between $\CVB$ and
$\CW$ is possible under AD as long as the two states being compared
occupy the same excitation sector} -- irrespective of $N$ and
irrespective of whether that sector eventually undergoes ESD. 
The vertex-base link's same-sector proportional advantage is therefore
preserved under amplitude damping throughout the interval in which
the corresponding concurrences remain nonzero.

\begin{figure}[t!]
\centering
\includegraphics[width=0.5\textwidth]{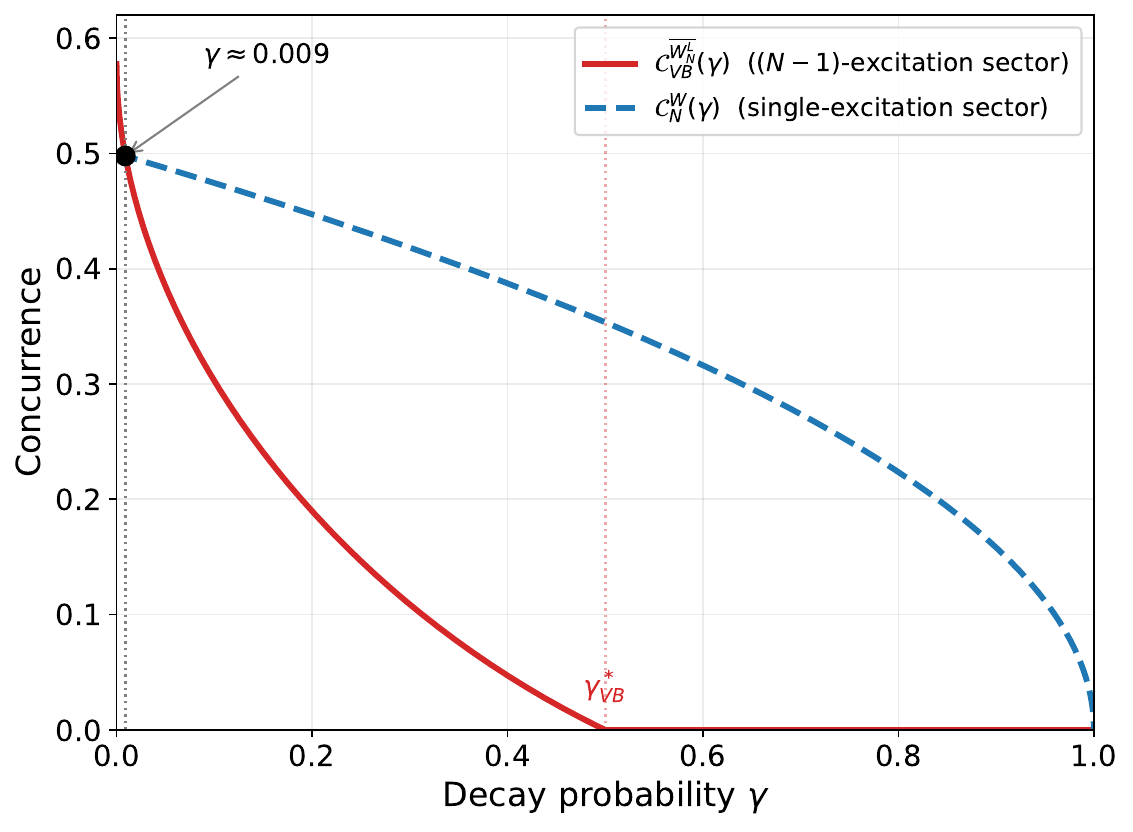}
\caption{Cross-sector comparison under AD at $N=4$: the
	$(N-1)$-excitation vertex-base concurrence
	$\mathcal{C}_{VB}^{\overline{W_N^L}}(\gamma)$ (red) against the
	single-excitation symmetric concurrence $\mathcal{C}^{W}_{N}(\gamma)$
	(blue). The former starts above the latter but reaches ESD at
	$\gamma^{*}_{VB}=1/(N-2)$ (dotted vertical line), whereas the
	single-excitation symmetric concurrence remains nonzero for
	$\gamma<1$. A crossing is therefore unavoidable for this particular
	comparison. The crossing (black dot) occurs at
	$\gamma\approx0.009$, showing how the initial vertex-base concurrence
	advantage can be reversed when the compared states belong to
	excitation sectors with different AD responses.}
\label{fig:crosssector}
\end{figure}

The three-qubit comparison made in our recent work~\cite{bhattacharyya2026super} was cross-sector:
the two-excitation Lohmayer state $|W_{3}^{L}\rangle$ was compared
with the conventional single-excitation symmetric state $|W_{3}\rangle$.
Under AD, the former reaches zero concurrence at a finite noise
strength whereas the latter remains entangled for every
$\gamma<1$. Since the Lohmayer vertex-base concurrence is initially
larger than the symmetric concurrence, continuity then guarantees
a crossing between these two particular curves. The present
framework shows that this crossing results from combining the
initial vertex-base concurrence advantage with different
excitation-sector responses to AD and therefore does not, by itself,
establish an intrinsic fragility of the vertex-base geometry. At $N=4$, the analogous cross-sector crossing occurs already at $\gamma\approx0.009$. Figure~\ref{fig:crosssector} illustrates this cross-sector reordering
for $N=4$.

\subsection{What actually reorders, and what does not}

Separating the two axes gives a cleaner picture than the original
single hierarchy comparison:

\textbf{The same-sector VB--W ordering does not invert for the channels studied here.}
Within either excitation sector, the vertex-base concurrence retains
its proportional advantage over the symmetric reference throughout
the range in which both concurrences remain nonzero, under PD, AD,
DP, and GAD (compare Section~\ref{sec:asym-dyn}).

\begin{table*}[t!]
	\centering
	\renewcommand{\arraystretch}{1.4}
	\resizebox{\textwidth}{!}{%
		\begin{tabular}{@{}llll@{}}
			\toprule
			\textbf{Channel} &
			\textbf{Symmetric family} &
			\textbf{Asymmetric family} &
			\textbf{Notes} \\
			\midrule
			
			PD &
			$\mathcal{C}_{W}$ decays as $2\sqrt{1-p}/N$ &
			$\mathcal{C}_{VB},\mathcal{C}_{BB}$ decay as $\sqrt{1-p}$ times initial &
			$\mathcal{C}_{VB}>\mathcal{C}_{W}>\mathcal{C}_{BB}$ for $0\le p<1$; no ESD
			\\
			
			AD &
			$\bar{W}_N$: $\gamma^{*}=1/(N-2)$; ESD for $N\ge4$ &
			$W^L_b$: $\gamma^{*}_{VB}=1/(N-2)$,
			$\gamma^{*}_{BB}=1/[2(N-2)]$ &
			At $N=3$, only BB has ESD; for $N\ge4$, BB has the lower asymmetric pair-class threshold
			\\
			
			DP &
			Common $p^{*}=3/[2(N+1)]$ &
			$p^{*}_{VB}=3/[2(N+1)]$,
			$p^{*}_{BB}=3/[2(2N-1)]$ &
			Bit-flip symmetry restored
			\\
			
			GAD &
			$\mathcal{C}^{W}_{N}(p,\alpha)=
			\mathcal{C}^{\bar W}_{N}(p,1-\alpha)$ &
			$\mathcal{C}^{W^{L}}_{VB,BB}(p,\alpha)=
			\mathcal{C}^{\overline{W^{L}}}_{VB,BB}(p,1-\alpha)$ &
			Bit-flip covariance; identical dynamics at
			$\alpha^{*}=\frac{1}{2}$
			\\
			
			\bottomrule
		\end{tabular}%
	}
	
	\caption{Comparative summary of pairwise entanglement dynamics for
		symmetric and asymmetric $N$-qubit $|W\rangle$-class states under
		one-sided local noise.}
	\label{tab:full-summary}
\end{table*}

\textbf{Cross-sector comparisons can reorder under AD.}
For the comparison considered in our recent work~\cite{bhattacharyya2026super}, the
$(N-1)$-excitation vertex-base concurrence starts above the
single-excitation symmetric reference but reaches ESD at finite
noise strength, whereas the latter remains entangled for
$\gamma<1$. A crossing is therefore unavoidable in this case.
The reordering reflects the different AD response of the two
excitation sectors and therefore does not, by itself, establish an
intrinsic fragility of the vertex-base geometry.

\textbf{The base-base link shows a distinct same-sector structural fragility.} Unlike the VB-vs-$W$ comparison, the lower
robustness of the BB link is not an artifact of sector mismatch.
Under AD, within the same $(N-1)$-excitation sector,
$\gamma^{*}_{BB}=\tfrac{1}{2}\gamma^{*}_{VB}$ for every $N\ge3$,
with $\gamma^{*}_{VB}=1$ lying at the physical boundary when $N=3$.
Under DP, $p^{*}_{BB}<p^{*}_{VB}$ holds within either excitation
sector for every $N\ge3$. This is a same-sector structural effect
arising from the distinct population structures of the BB and VB
reductions (Appendices~\ref{app:noiseE-VB} and~\ref{app:noiseF-VB}).
Under AD, the relevant excited-state population in the BB reduction
is exactly twice that of the VB reduction, yielding
$\gamma^{*}_{BB}=\tfrac{1}{2}\gamma^{*}_{VB}$; under DP, the
different population terms likewise give
$p^{*}_{BB}<p^{*}_{VB}$.

The three-qubit analysis combined two distinct effects under the
term ``super-link fragility'': a cross-sector reordering involving
the vertex-base link, and the lower ESD threshold of the peripheral
base-base link. The present framework distinguishes these mechanisms.
We therefore associate the same-sector structural fragility with
the base-base pair class, whose entanglement is lost earlier than that
of the vertex-base pair class under DP and, for the $(N-1)$-excitation
states, under AD, while treating the cross-sector VB--$W$ crossing as
an excitation-sector effect.
In this refined picture, the vertex-base link retains its proportional
advantage over the symmetric reference within a fixed excitation sector
for all values at which both concurrences are nonzero, whereas the
base-base link exhibits a distinct same-sector structural fragility.

\section{Discussion}
\label{sec:disc}

Combining Sections~\ref{sec:sym-dyn}, \ref{sec:asym-dyn} and
\ref{sec:reinterp}, the pairwise entanglement dynamics considered
here can be organized in terms of three distinct factors: network
geometry, excitation sector, and noise symmetry.

\paragraph{Network geometry\\}
The symmetric family realises an equilateral weighted network with
uniform initial pairwise concurrence $2/N$, whereas the asymmetric
family realises a star-plus-clique weighted network with two
inequivalent pair classes having initial concurrences
$\CVB=1/\sqrt{N-1}$ and $\CBB=1/(N-1)$, satisfying
$\CVB>\CW>\CBB$.
This structural asymmetry produces distinct
vertex-base and base-base reduced states and, relative to the
symmetric reference, fixes the initial ordering
$\CVB>\CW>\CBB$. Under local noise, the inequivalent VB
and BB reduced-state structures produce link-dependent dynamics, most
clearly seen in the lower BB ESD threshold under DP and, within the
$(N-1)$-excitation sector, under AD.

\paragraph{Excitation sector\\}
Under AD, the single-excitation states are ESD-free for
$\gamma<1$, whereas their $(N-1)$-excitation partners can exhibit ESD. For $N\ge4$, the symmetric $\WbarN$ state and the
vertex-base link of $\WLb$ undergo ESD at
$\gamma^{*}=1/(N-2)$, while the base-base link of $\WLb$ has the
lower threshold $\gamma^{*}_{BB}=1/[2(N-2)]$. At $N=3$, the first
two thresholds lie at the boundary $\gamma=1$, while the BB link
alone exhibits ESD.

\paragraph{Noise symmetry\\}
PD modifies only the coherence term while leaving the diagonal
populations unchanged; consequently, it never activates the
$\sqrt{ad}$ penalty and never induces ESD.
AD is directional: it distinguishes bit-flipped partners because
relaxation acts differently on the single- and $(N-1)$-excitation
sectors. The $(N-1)$-excitation partners can therefore exhibit ESD, whereas the corresponding single-excitation states
remain ESD-free for $\gamma<1$.
DP is bit-flip symmetric: within each pair class, the two globally
bit-flipped partners have identical concurrence dynamics and therefore
the same ESD threshold.
GAD interpolates between relaxation- and excitation-dominated
regimes: the substitution $\alpha\to1-\alpha$ maps each state to its
bit-flipped partner, with identical concurrence dynamics at the
symmetry point $\alpha^{*}=\tfrac{1}{2}$.

Table~\ref{tab:full-summary} summarizes the channel-by-channel
comparison of the symmetric and asymmetric families.

\section{Conclusions}
\label{sec:conc}

We have introduced an analytically tractable $N$-qubit generalization of the asymmetric Lohmayer $W$-class geometry and derived closed-form pairwise concurrence dynamics for both the symmetric and asymmetric families under representative one-sided noise models. The construction separates two previously intertwined ingredients: network geometry, through the distinction between vertex-base and base-base pair classes, and excitation sector, through the single- and $(N-1)$-excitation partners.

A central result is that, within a fixed excitation sector, the vertex-base concurrence retains the same noise dependence as the corresponding symmetric reference and therefore preserves its proportional advantage throughout the range in which both concurrences remain nonzero. The amplitude-damping reordering identified previously for the three-qubit Lohmayer state is consequently a cross-sector effect: it arises because states with different excitation content respond differently to directional relaxation, rather than because the vertex-base geometry is intrinsically fragile. By contrast, the base-base pair exhibits a distinct same-sector structural fragility, manifested by its lower entanglement-sudden-death threshold relative to the vertex-base pair under depolarizing noise and, for the $(N-1)$-excitation family, under amplitude damping.

These results show that pairwise entanglement robustness in asymmetric multipartite networks cannot be inferred from the initial concurrence of a link alone. Its dynamics depend jointly on the weighted network structure, the excitation sector of the underlying state, and the symmetry of the noise process. Separating these ingredients provides a systematic way to distinguish genuine structural fragility from apparent reordering caused by comparing inequivalent excitation sectors, and offers a framework for analyzing entanglement distribution in more general asymmetric quantum networks.

\acknowledgements{
F.O. acknowledges financial support from Tokyo International University Personal Research Fund and Special
Grant-in-Aid for Research Work. }

\onecolumngrid
\newpage

\appendix

\section{Supporting Results and Figures}
\label{app:supporting}

\subsection{Symmetric-family GAD dynamics}

\begin{figure}[H]
	\centering
	\includegraphics[width=0.9\columnwidth]{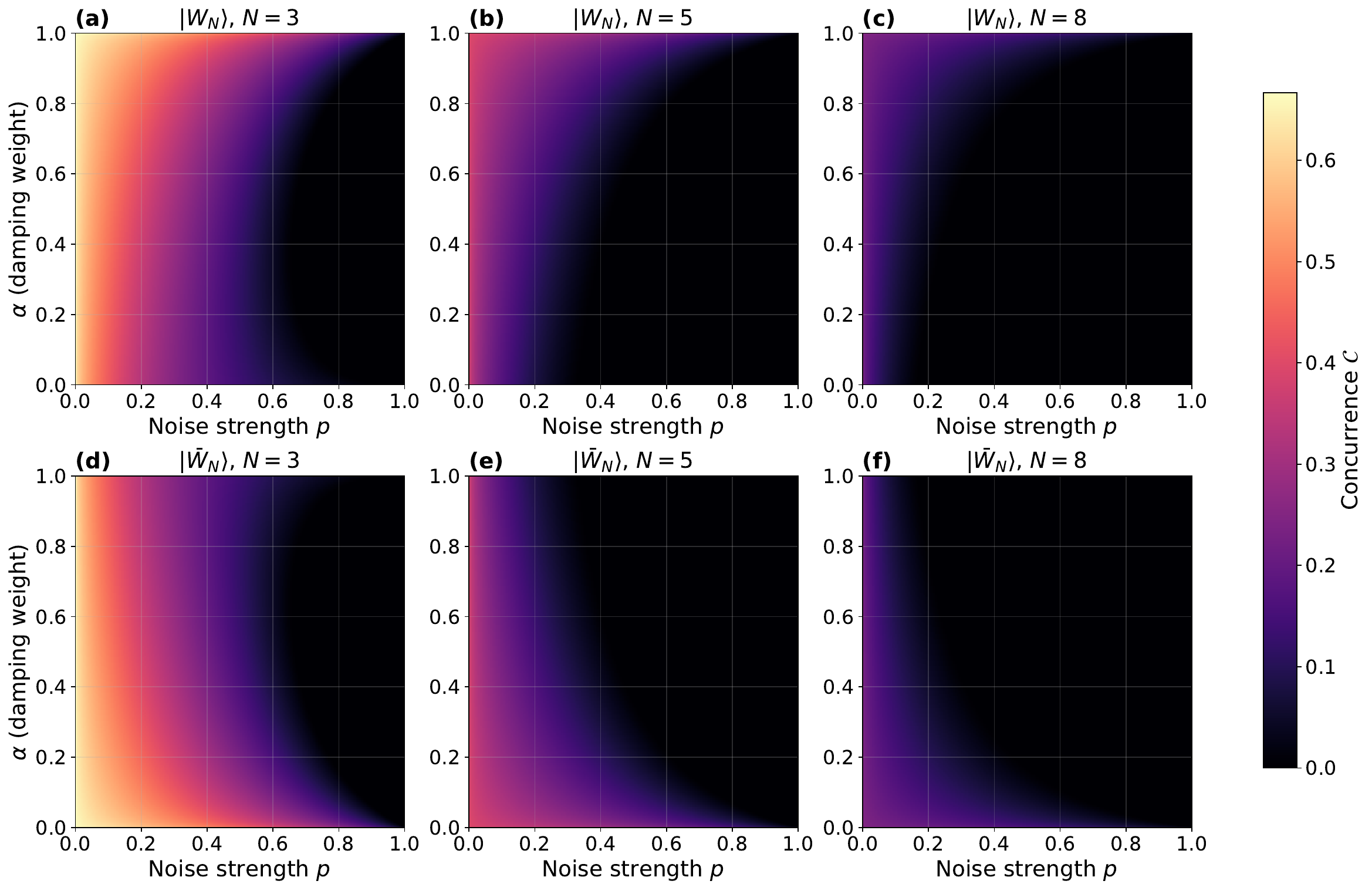}
	\caption{GAD concurrence heatmaps for $N=3$, $5$, and $8$ (columns) for the symmetric family, showing $\WN$ (top row) and $\WbarN$ (bottom row). The two rows are related by $\alpha\!\leftrightarrow\!1-\alpha$.}
	\label{fig:GADsymm}
\end{figure}

Figure~\ref{fig:GADsymm} visualizes the GAD concurrence dynamics and
the bit-flip covariance discussed in Section~\ref{sec:sym-dyn}.

\subsection{Residual coherence and vertex signature}

\begin{figure}[H]
	\centering
	\includegraphics[width=0.5\textwidth]{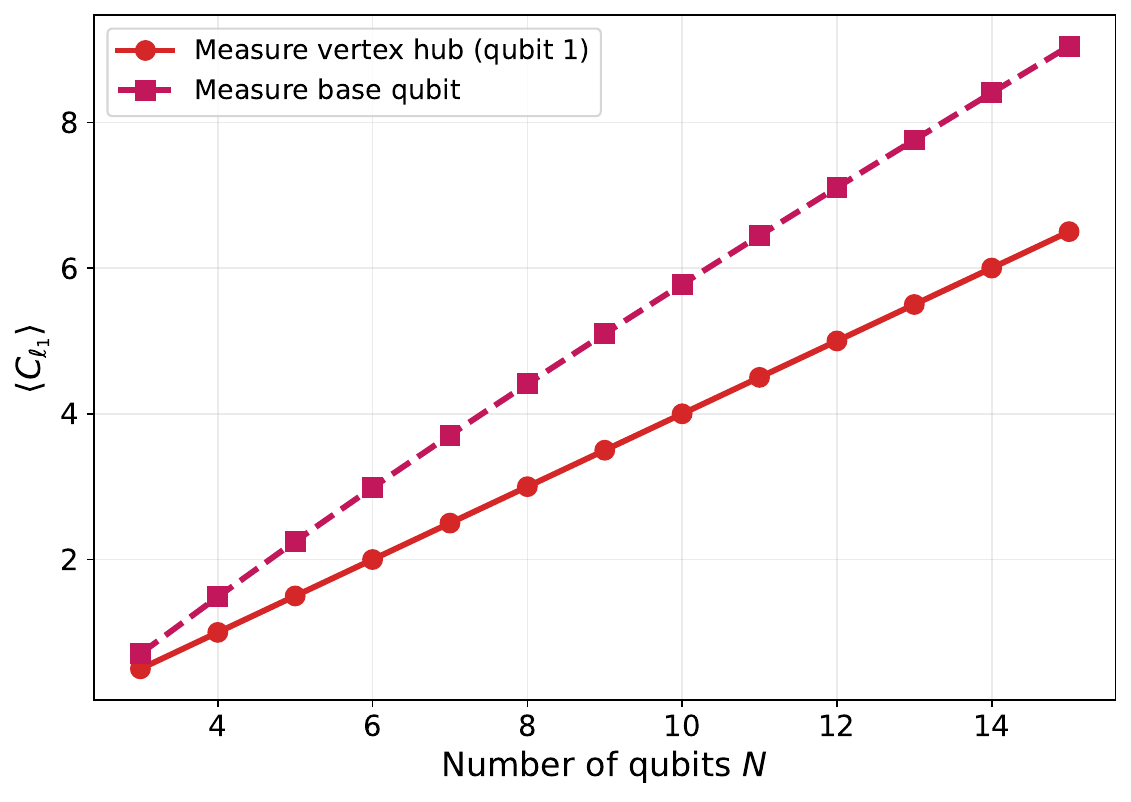}
	\caption{Average residual $\ell_{1}$-coherence after a local
		measurement on the vertex (solid) versus on a base qubit (dashed).
		The gap widens with $N$, providing a structural signature of the
		distinguished vertex.}
	\label{fig:residcoh}
\end{figure}

The average residual $\ell_{1}$-coherence after a local projective
measurement of $\WL$ depends on which qubit is measured
(Appendix~\ref{app:initial-VB}):
\begin{align}
	\langle\Cl\rangle_{V}^{W^{L}} &=\frac{N-2}{2},
	\label{eq:CresV}\\
	\langle\Cl\rangle_{B}^{W^{L}}
	&=\frac{(N-2)\bigl[(N-3)+2\sqrt{N-1}\bigr]}{2(N-1)}.
	\label{eq:CresB}
\end{align}
Both results are identical for $\WLb$ by bit-flip symmetry. The
difference $\langle\Cl\rangle_{B}-\langle\Cl\rangle_{V}>0$ for all
$N\ge3$ provides a coherence-level signature of the structural
asymmetry: measuring a base qubit leaves a more coherent residual
state on average than measuring the vertex.
Together with the concurrence hierarchy of Eq.~\eqref{eq:hierarchy}, this provides
a complementary structural signature of the distinguished vertex qubit.
Figure~\ref{fig:residcoh} illustrates the resulting vertex--base
difference in average residual coherence as a function of $N$.

\subsection{Asymmetric-family supporting figures}

\paragraph{Phase damping.}

\begin{figure}[H]
	\centering
	\includegraphics[width=0.5\columnwidth]{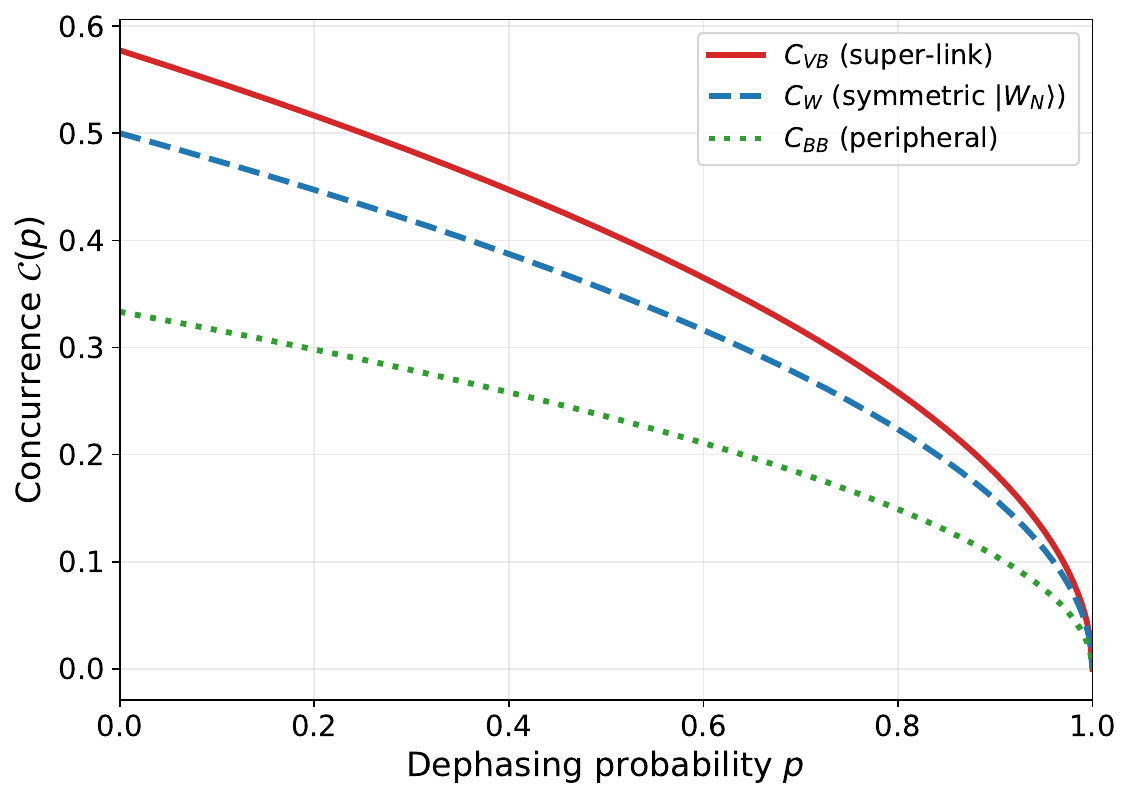}
	\caption{PD concurrence for the asymmetric and symmetric $N$-qubit
		$|W\rangle$-class states ($N=4$). The hierarchy
		$\CVB>\CW>\CBB$ is preserved for $0\le p<1$, and no ESD occurs.}
	\label{fig:PDasym}
\end{figure}

Figure~\ref{fig:PDasym} illustrates the phase-damping concurrence
dynamics and the preserved concurrence hierarchy discussed in
Section~\ref{sec:asym-dyn}.

\paragraph{GAD heatmaps.}

\begin{figure}[H]
	\centering
	\includegraphics[width=0.9\columnwidth]{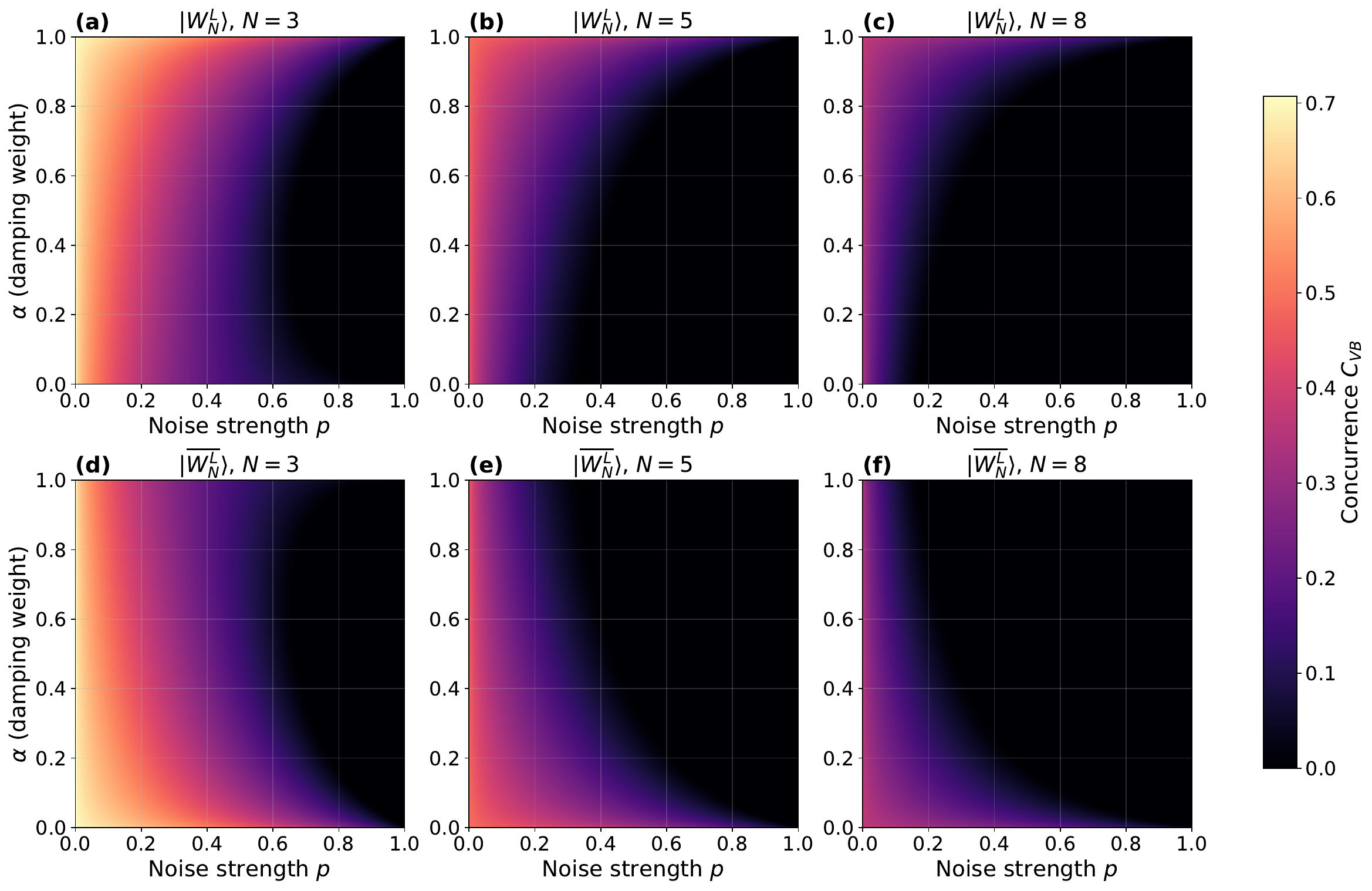}
	\caption{VB concurrence heatmaps in $(p,\alpha)$ for $N=3,5,8$.
		\emph{Top:} $\WL$. \emph{Bottom:} $\WLb$. The two rows are related
		by $\alpha\!\leftrightarrow\!1-\alpha$.}
	\label{fig:GADasymVBhm}
\end{figure}

Figure~\ref{fig:GADasymVBhm} shows the vertex-base concurrence
landscape in the $(p,\alpha)$ plane for representative system sizes
and illustrates the $\alpha\!\leftrightarrow\!1-\alpha$ covariance.

\begin{figure}[H]
	\centering
	\includegraphics[width=0.8\columnwidth]{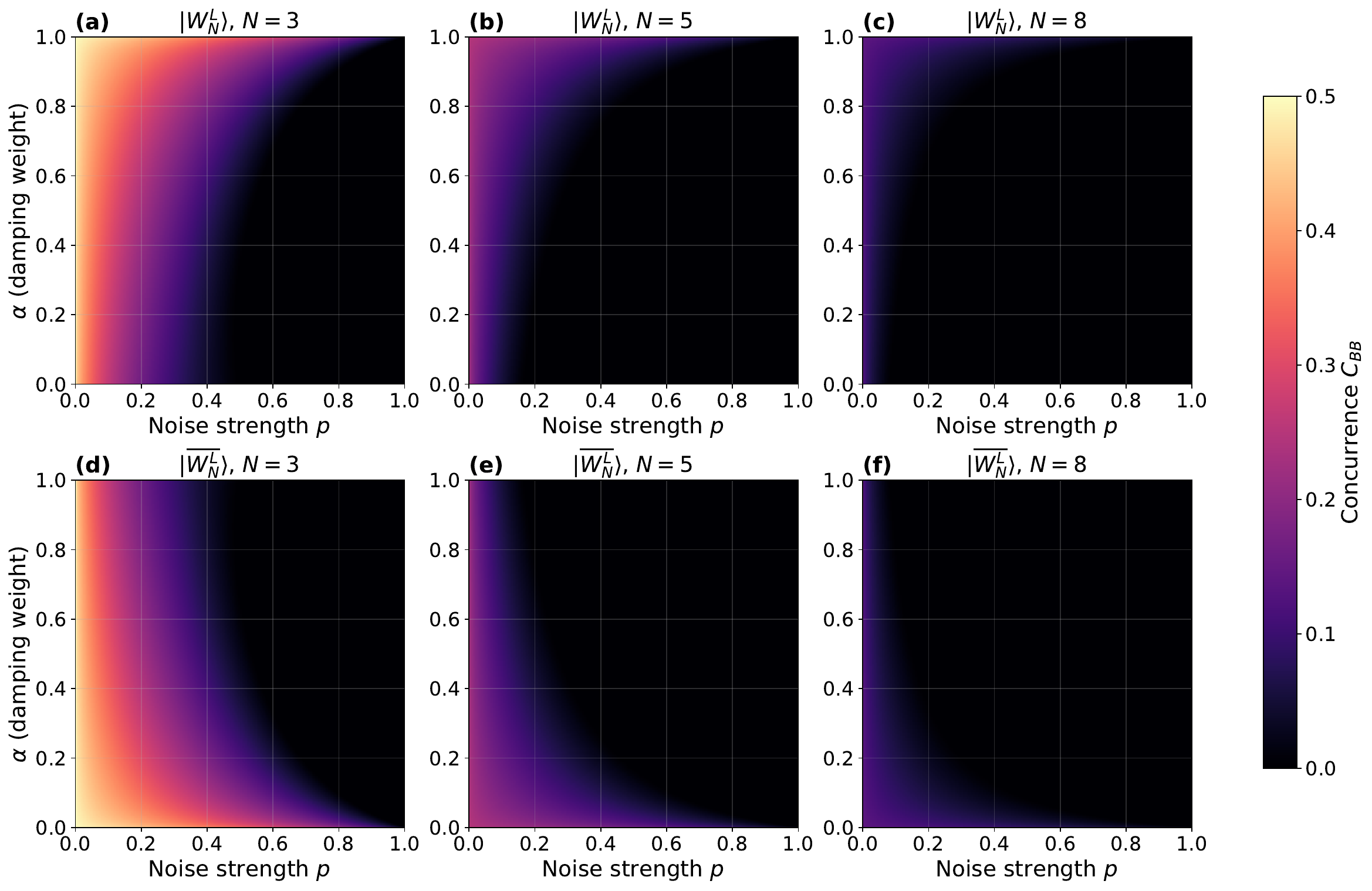}
	\caption{BB concurrence heatmaps in $(p,\alpha)$ for $N=3,5,8$. For
		each $N$, the BB entangled region is contained within the corresponding
		VB region and is strictly smaller overall.}
	\label{fig:GADasymBBhm}
\end{figure}

Figure~\ref{fig:GADasymBBhm} shows the base-base concurrence
landscape in the $(p,\alpha)$ plane and illustrates the containment
of the BB entangled region within the corresponding VB region.

\section{General $X$-State Transformations under One-Sided Noise}
\label{app:general}

This appendix collects the channel transformations used throughout
the subsequent derivations. Starting from a generic two-qubit
$X$ state, we determine how the matrix elements relevant to the
concurrence evolve under the one-sided noise convention of the main
text. The resulting formulas provide a common starting point for
Appendices~\ref{app:noiseC-W}--\ref{app:noiseF-VB}.

\subsection{Setup}
\begin{equation}
\rho_X =
\begin{pmatrix}
a & 0 & 0 & w\\
0 & b & z & 0\\
0 & z^{*} & c & 0\\
w^{*} & 0 & 0 & d
\end{pmatrix},
\qquad a+b+c+d=1,\quad a,b,c,d\ge0,\ z,w\in\mathbb{C}.
\end{equation}
Every channel below acts as $E_k=I\otimes K_k$ (single-qubit noise on the \emph{second} qubit
of whichever pair is being considered), and
\begin{equation}
\rho_X' = \sum_k E_k\,\rho_X\,E_k^\dagger.
\end{equation}

\textbf{Notation:}
Unprimed $(a,b,c,d,z,w)$ denotes the \emph{input} state $\rho_X$, before the channel acts.
Primed $(a',b',c',d',z',w')$ denotes the \emph{output} state $\rho_X'$, after the channel
acts. 
Each ``Sum'' formula below gives the transformed matrix elements in
terms of the input parameters; the final subsection then evaluates
the concurrence directly from the resulting primed entries.\\

\textbf{Second-qubit convention:} for a vertex-base pair $\rho_{12}$, noise acts on qubit~2
(base); for a base-base pair $\rho_{23}$, noise acts on qubit~3. Both are qubit-2-of-the-pair,
so the formulas below apply unchanged to either.

\subsection{Phase Damping}

\subsubsection{Kraus Operators}
\begin{align}
K_0 =\begin{pmatrix}1&0\\0&\sqrt{1-p}\end{pmatrix},\qquad
K_1=\begin{pmatrix}0&0\\0&\sqrt{p}\end{pmatrix}.
\end{align}
\begin{align}
E_0=I\otimes K_0=\begin{pmatrix}1&0&0&0\\0&\sqrt{1-p}&0&0\\0&0&1&0\\0&0&0&\sqrt{1-p}\end{pmatrix},
\qquad
E_1=I\otimes K_1=\begin{pmatrix}0&0&0&0\\0&\sqrt{p}&0&0\\0&0&0&0\\0&0&0&\sqrt{p}\end{pmatrix}.
\end{align}
Both are real diagonal, so $E_k^\dagger=E_k$.

\subsubsection{Term $k=0$}
\begin{equation}
E_0\,\rho_X\,E_0^\dagger=
\begin{pmatrix}
a & 0 & 0 & w\sqrt{1-p}\\
0 & b(1-p) & z\sqrt{1-p} & 0\\
0 & z^{*}\sqrt{1-p} & c & 0\\
w^{*}\sqrt{1-p} & 0 & 0 & d(1-p)
\end{pmatrix}.
\end{equation}

\subsubsection{Term $k=1$}
\begin{equation}
E_1\,\rho_X\,E_1^\dagger=
\begin{pmatrix}
0&0&0&0\\ 0& bp &0&0\\ 0&0&0&0\\ 0&0&0& dp
\end{pmatrix}.
\end{equation}

\subsubsection{Sum}
\begin{equation}
\rho_X'=
\begin{pmatrix}
a & 0 & 0 & w\sqrt{1-p}\\
0 & b & z\sqrt{1-p} & 0\\
0 & z^{*}\sqrt{1-p} & c & 0\\
w^{*}\sqrt{1-p} & 0 & 0 & d
\end{pmatrix}
\end{equation}
Trace check: $a+b+c+d=1$.

\subsection{Amplitude Damping}

\subsubsection{Kraus Operators}
\begin{align}
K_0=\begin{pmatrix}1&0\\0&\sqrt{1-\gamma}\end{pmatrix},\qquad
K_1=\begin{pmatrix}0&\sqrt{\gamma}\\0&0\end{pmatrix}.
\end{align}
\begin{align}
E_0=\begin{pmatrix}1&0&0&0\\0&\sqrt{1-\gamma}&0&0\\0&0&1&0\\0&0&0&\sqrt{1-\gamma}\end{pmatrix},
\qquad
E_1=\begin{pmatrix}0&\sqrt{\gamma}&0&0\\0&0&0&0\\0&0&0&\sqrt{\gamma}\\0&0&0&0\end{pmatrix}.
\end{align}
$E_0^\dagger=E_0$; $E_1^\dagger\ne E_1$.

\subsubsection{Term $k=0$}
\begin{equation}
E_0\,\rho_X\,E_0^\dagger=
\begin{pmatrix}
a & 0 & 0 & w\sqrt{1-\gamma}\\
0 & b(1-\gamma) & z\sqrt{1-\gamma} & 0\\
0 & z^{*}\sqrt{1-\gamma} & c & 0\\
w^{*}\sqrt{1-\gamma} & 0 & 0 & d(1-\gamma)
\end{pmatrix}.
\end{equation}

\subsubsection{Term $k=1$}
\begin{equation}
E_1\,\rho_X\,E_1^\dagger=
\begin{pmatrix}
b\gamma&0&0&0\\ 0&0&0&0\\ 0&0& d\gamma &0\\ 0&0&0&0
\end{pmatrix}.
\end{equation}

\subsubsection{Sum}
\begin{equation}
\rho_X'=
\begin{pmatrix}
a+b\gamma & 0 & 0 & w\sqrt{1-\gamma}\\
0 & b(1-\gamma) & z\sqrt{1-\gamma} & 0\\
0 & z^{*}\sqrt{1-\gamma} & c+d\gamma & 0\\
w^{*}\sqrt{1-\gamma} & 0 & 0 & d(1-\gamma)
\end{pmatrix}
\end{equation}
Trace check: $(a+b\gamma)+b(1-\gamma)+(c+d\gamma)+d(1-\gamma)=a+b+c+d=1$.

\subsection{Depolarization}

\subsubsection{Kraus Operators}
\begin{align}
K_0=\sqrt{1-p}\,I,\quad K_1=\sqrt{\tfrac p3}\,\sigma_x,\quad K_2=\sqrt{\tfrac p3}\,\sigma_y,\quad K_3=\sqrt{\tfrac p3}\,\sigma_z.
\end{align}
\begin{align}
E_0&=\sqrt{1-p}\,I_4, &
E_1&=\sqrt{\tfrac p3}\begin{pmatrix}0&1&0&0\\1&0&0&0\\0&0&0&1\\0&0&1&0\end{pmatrix},\\[4pt]
E_2&=\sqrt{\tfrac p3}\begin{pmatrix}0&-i&0&0\\i&0&0&0\\0&0&0&-i\\0&0&i&0\end{pmatrix}, &
E_3&=\sqrt{\tfrac p3}\begin{pmatrix}1&0&0&0\\0&-1&0&0\\0&0&1&0\\0&0&0&-1\end{pmatrix}.
\end{align}
All $E_k$ Hermitian.

\subsubsection{Term $k=0$ (identity)}
\begin{equation}
E_0\,\rho_X\,E_0^\dagger=(1-p)\,\rho_X=(1-p)
\begin{pmatrix}
a&0&0&w\\0&b&z&0\\0&z^{*}&c&0\\w^{*}&0&0&d
\end{pmatrix}.
\end{equation}

\subsubsection{Term $k=1$ ($\sigma_x$ on qubit 2)}
\begin{equation}
E_1\,\rho_X\,E_1^\dagger=\frac p3
\begin{pmatrix}
b&0&0&z\\0&a&w&0\\0&w^{*}&d&0\\z^{*}&0&0&c
\end{pmatrix}.
\end{equation}

\subsubsection{Term $k=2$ ($\sigma_y$ on qubit 2)}
\begin{equation}
E_2\,\rho_X\,E_2^\dagger=\frac p3
\begin{pmatrix}
b&0&0&-z\\0&a&-w&0\\0&-w^{*}&d&0\\-z^{*}&0&0&c
\end{pmatrix}.
\end{equation}

\subsubsection{Term $k=3$ ($\sigma_z$ on qubit 2)}
\begin{equation}
E_3\,\rho_X\,E_3^\dagger=\frac p3
\begin{pmatrix}
a&0&0&-w\\0&b&-z&0\\0&-z^{*}&c&0\\-w^{*}&0&0&d
\end{pmatrix}.
\end{equation}

\subsubsection{Sum}
Adding the last three terms first:
\begin{equation}
E_1(\cdot)E_1^\dagger+E_2(\cdot)E_2^\dagger+E_3(\cdot)E_3^\dagger
=\frac p3
\begin{pmatrix}
a+2b&0&0&-w\\0&2a+b&-z&0\\0&-z^{*}&2d+c&0\\-w^{*}&0&0&2c+d
\end{pmatrix},
\end{equation}
then adding $(1-p)\rho_X$ and collecting terms:
\begin{equation}
\rho_X'=
\begin{pmatrix}
a+\tfrac{2p}{3}(b-a) & 0 & 0 & w\bigl(1-\tfrac{4p}{3}\bigr)\\[4pt]
0 & b+\tfrac{2p}{3}(a-b) & z\bigl(1-\tfrac{4p}{3}\bigr) & 0\\[4pt]
0 & z^{*}\bigl(1-\tfrac{4p}{3}\bigr) & c+\tfrac{2p}{3}(d-c) & 0\\[4pt]
w^{*}\bigl(1-\tfrac{4p}{3}\bigr) & 0 & 0 & d+\tfrac{2p}{3}(c-d)
\end{pmatrix}
\end{equation}
Trace check: sum of diagonal $=a+b+c+d=1$.

\subsection{Generalized Amplitude Damping Channel (GAD)}
\label{sec:gadc}

A single GAD convention is used throughout this document for all four
states, $\WN$, $\WbarN$, $\WL$, and $\WLb$. This matches the Kraus operators
of Eq.~\eqref{eq:kraus-GAD} in the main text and the convention used in our recent
three-qubit work~\cite{bhattacharyya2026super}.

\subsubsection{Kraus Operators}
\begin{align}
K_0&=\sqrt{\alpha}\begin{pmatrix}1&0\\0&\sqrt{1-p}\end{pmatrix}, &
K_1&=\sqrt{\alpha}\begin{pmatrix}0&\sqrt{p}\\0&0\end{pmatrix},\\[4pt]
K_2&=\sqrt{1-\alpha}\begin{pmatrix}\sqrt{1-p}&0\\0&1\end{pmatrix}, &
K_3&=\sqrt{1-\alpha}\begin{pmatrix}0&0\\\sqrt{p}&0\end{pmatrix}.
\end{align}
Parameters: $p\in[0,1]$ (damping probability), $\alpha\in[0,1]$ (thermal weight).
$\alpha=1$ reduces to ordinary amplitude damping; $\alpha=0$ is the pure-excitation
(pumping) limit.

\begin{align}
E_0&=\sqrt{\alpha}\begin{pmatrix}1&0&0&0\\0&\sqrt{1-p}&0&0\\0&0&1&0\\0&0&0&\sqrt{1-p}\end{pmatrix}, &
E_1&=\sqrt{\alpha}\begin{pmatrix}0&\sqrt{p}&0&0\\0&0&0&0\\0&0&0&\sqrt{p}\\0&0&0&0\end{pmatrix},\\[6pt]
E_2&=\sqrt{1-\alpha}\begin{pmatrix}\sqrt{1-p}&0&0&0\\0&1&0&0\\0&0&\sqrt{1-p}&0\\0&0&0&1\end{pmatrix}, &
E_3&=\sqrt{1-\alpha}\begin{pmatrix}0&0&0&0\\\sqrt{p}&0&0&0\\0&0&0&0\\0&0&\sqrt{p}&0\end{pmatrix}.
\end{align}

\subsubsection{Term $k=0$}
\begin{equation}
E_0\,\rho_X\,E_0^\dagger=\alpha
\begin{pmatrix}
a&0&0&w\sqrt{1-p}\\0&b(1-p)&z\sqrt{1-p}&0\\0&z^{*}\sqrt{1-p}&c&0\\w^{*}\sqrt{1-p}&0&0&d(1-p)
\end{pmatrix}.
\end{equation}

\subsubsection{Term $k=1$}
\begin{equation}
E_1\,\rho_X\,E_1^\dagger=\alpha p
\begin{pmatrix}
b&0&0&0\\0&0&0&0\\0&0&d&0\\0&0&0&0
\end{pmatrix}.
\end{equation}

\subsubsection{Term $k=2$}
\begin{equation}
E_2\,\rho_X\,E_2^\dagger=(1-\alpha)
\begin{pmatrix}
a(1-p)&0&0&w\sqrt{1-p}\\0&b&z\sqrt{1-p}&0\\0&z^{*}\sqrt{1-p}&c(1-p)&0\\w^{*}\sqrt{1-p}&0&0&d
\end{pmatrix}.
\end{equation}

\subsubsection{Term $k=3$}
\begin{equation}
E_3\,\rho_X\,E_3^\dagger=(1-\alpha)p
\begin{pmatrix}
0&0&0&0\\0&a&0&0\\0&0&0&0\\0&0&0&c
\end{pmatrix}.
\end{equation}

\subsubsection{Sum}

\begin{equation}
\rho_X'=
\begin{pmatrix}
a' & 0 & 0 & w\sqrt{1-p}\\
0 & b' & z\sqrt{1-p} & 0\\
0 & z^{*}\sqrt{1-p} & c' & 0\\
w^{*}\sqrt{1-p} & 0 & 0 & d'
\end{pmatrix}
\end{equation}
with
\begin{align}
a'&=a(1-p)+\alpha p(a+b),\\
b'&=(1-\alpha)pa+b(1-\alpha p),\\
c'&=c(1-p)+\alpha p(c+d),\\
d'&=(1-\alpha)pc+d(1-\alpha p).
\end{align}
Coherence law:
\begin{equation}
z'=z\sqrt{1-p},\qquad w'=w\sqrt{1-p}.
\end{equation}
Trace check: $a'+b'+c'+d' = (a+b)+(c+d)=1$.\\

\noindent\textbf{Key property:} the off-diagonal coherence under this convention decays as
$\sqrt{1-p}$, \emph{independently of $\alpha$}. All $\alpha$-dependence enters only
through the diagonal populations $a',d'$.

\subsection{State-Specific Parameters}
\label{sec:plugin}
\begin{center}
\renewcommand{\arraystretch}{2} 
\begin{tabular}{|l|c|c|c|c|c|c|}
\hline
State & $a$ & $b$ & $c$ & $d$ & $z$ & $w$\\
\hline
$\rho_{12}^{W}$ & $\frac{N-2}N$ & $\frac1N$ & $\frac1N$ & $0$ & $\frac1N$ & $0$\\
$\rho_{12}^{\bar W}$ & $0$ & $\frac1N$ & $\frac1N$ & $\frac{N-2}N$ & $\frac1N$ & $0$\\
\hline
$\rho_{12}^{W^{L}}$ (VB) & $\frac{N-2}{2(N-1)}$ & $\frac{1}{2(N-1)}$ & $\frac12$ & $0$ & $\frac1{2\sqrt{N-1}}$ & $0$\\
$\rho_{23}^{W^{L}}$ (BB) & $\frac{N-2}{N-1}$ & $\frac1{2(N-1)}$ & $\frac1{2(N-1)}$ & $0$ & $\frac1{2(N-1)}$ & $0$\\
$\rho_{12}^{\overline{W^{L}}}$ (VB) & $0$ & $\frac12$ & $\frac1{2(N-1)}$ & $\frac{N-2}{2(N-1)}$ & $\frac1{2\sqrt{N-1}}$ & $0$\\
$\rho_{23}^{\overline{W^{L}}}$ (BB) & $0$ & $\frac1{2(N-1)}$ & $\frac1{2(N-1)}$ & $\frac{N-2}{N-1}$ & $\frac1{2(N-1)}$ & $0$\\
\hline
\end{tabular}
\renewcommand{\arraystretch}{1} 
\end{center}
The table collects the $X$-state parameters required for all
subsequent calculations. Substitution into the channel
transformations derived above yields the state-specific results of
Appendices~\ref{app:noiseC-W}--\ref{app:noiseF-VB}.

\subsection{Concurrence after One-Sided Noise}
\label{app:concurrence}
For any X-state $\rho_X$ as parametrized above, the Wootters
concurrence is
\begin{equation}
\mathcal{C}(\rho_X) = 2\max\Bigl\{0,\ |z|-\sqrt{ad},\ |w|-\sqrt{bc}\Bigr\}
\end{equation}
Since every channel below maps X-states to X-states, this can be evaluated directly on
$(a',b',c',d',z',w')$ --- the reduced density matrix itself never needs to be written down.
Substituting the corresponding channel transformation gives:

\subsubsection{Phase Damping} ($a'=a,b'=b,c'=c,d'=d,\ z'=z\sqrt{1-p},\ w'=w\sqrt{1-p}$):
\begin{equation}
\mathcal{C} = 2\max\Bigl\{0,\ |z|\sqrt{1-p}-\sqrt{ad},\ |w|\sqrt{1-p}-\sqrt{bc}\Bigr\}
\end{equation}

\subsubsection{Amplitude Damping} ($a'=a+b\gamma,\ b'=b(1-\gamma),\ c'=c+d\gamma,\ d'=d(1-\gamma)$):
\begin{equation}
\mathcal{C} = 2\max\Bigl\{0,\
|z|\sqrt{1-\gamma}-\sqrt{(a+b\gamma)\,d(1-\gamma)},\
|w|\sqrt{1-\gamma}-\sqrt{b(1-\gamma)(c+d\gamma)}\Bigr\}
\end{equation}

\subsubsection{Depolarization} (with $a',b',c',d'$ given by the depolarization transformation above, $z'=z(1-\tfrac{4p}3)$, $w'=w(1-\tfrac{4p}3)$):
\begin{equation}
\mathcal{C} = 2\max\Bigl\{0,\ |z|\Bigl|1-\tfrac{4p}3\Bigr|-\sqrt{a'd'},\
|w|\Bigl|1-\tfrac{4p}3\Bigr|-\sqrt{b'c'}\Bigr\}
\end{equation}

\subsubsection{GAD} (with $a',b',c',d'$ as in Appendix~\ref{app:general}, $z'=z\sqrt{1-p}$, $w'=w\sqrt{1-p}$):
\begin{equation}
\mathcal{C} = 2\max\Bigl\{0,\ |z|\sqrt{1-p}-\sqrt{a'd'},\ |w|\sqrt{1-p}-\sqrt{b'c'}\Bigr\}
\end{equation}

For all six states in the table above, $w=0$, and each channel preserves $w'=0$. Hence the $|w'|-\sqrt{b'c'}$ branch vanishes identically, and the concurrence reduces to
\begin{equation}
\mathcal{C}=2\max\{0,\ |z'|-\sqrt{a'd'}\},
\end{equation}
which depends on $a',d',z'$ \emph{only} --- never on $b'$ or $c'$.


\section{Initial Properties of $|W_N\rangle$ and $|\bar{W}_N\rangle$}
\label{app:initial-W}

For completeness, we derive here the initial coherence, reduced
two-qubit states, and pairwise concurrence of the symmetric family.
These results establish the reference quantities used in the
same-sector comparisons with the asymmetric family in the main text.

\subsection{The Two States}

\begin{align}
|W_N\rangle &= \frac{1}{\sqrt{N}} \sum_{i=1}^{N} |0 \cdots 1_i \cdots 0\rangle, \\
|\bar{W}_N\rangle &= \frac{1}{\sqrt{N}} \sum_{i=1}^{N} |1 \cdots 0_i \cdots 1\rangle.
\end{align}

We note that, 
\begin{equation}
|\bar{W}_N\rangle = \sigma_x^{\otimes N} |W_N\rangle
\end{equation}

\subsection{$\ell_1$-Norm of Coherence}

For a pure state $|\psi\rangle = \sum_i c_i|i\rangle$,
\begin{equation}
\Cl(|\psi\rangle) = \Bigl(\sum_i |c_i|\Bigr)^2 - 1.
\end{equation}

For both $|W_N\rangle$ and $|\bar{W}_N\rangle$, each of the $N$ amplitudes has magnitude $1/\sqrt{N}$, so
\begin{equation}
\sum|c_i| = N \cdot \frac{1}{\sqrt{N}} = \sqrt{N}, \qquad
\Cl = N - 1.
\end{equation}

\subsubsection{After Measuring Qubit 1 (by permutation symmetry)}

\textbf{For $|W_N\rangle$:}
\begin{itemize}
\item Outcome $|1\rangle$ (prob.\ $1/N$): residual state $|0\cdots0\rangle$, coherence $=0$.
\item Outcome $|0\rangle$ (prob.\ $(N-1)/N$): residual state $|W_{N-1}\rangle$, coherence $=N-2$.
\end{itemize}
Average residual coherence:
\begin{equation}
\langle \Cl\rangle_{W_N} = \frac{(N-1)(N-2)}{N}.
\end{equation}

\textbf{For $|\bar{W}_N\rangle$:}
Roles of $|0\rangle,|1\rangle$ swap. Same probabilities, same average:
\begin{equation}
\langle \Cl\rangle_{\bar{W}_N} = \frac{(N-1)(N-2)}{N}.
\end{equation}

\subsection{Bipartite Reduced Density Matrices}

\subsubsection{For $|W_N\rangle$}

The density operator is
\begin{equation}
\rho_{W_N}=|W_N\rangle\langle W_N|=\frac1N\sum_{i,j=1}^N
|0\cdots1_i\cdots0\rangle\langle0\cdots1_j\cdots0|.
\end{equation}
Tracing out qubits $3,\ldots,N$: a term $|0\cdots1_i\cdots0\rangle\langle0\cdots1_j\cdots0|$ 
survives the partial trace only when qubits $3,\ldots,N$ agree in the bra and ket.

\paragraph{Diagonal contributions ($i=j$)\\}
\begin{itemize}
\item $i=1$: contributes $|10\rangle\langle10|$ with coefficient $1/N$.
\item $i=2$: contributes $|01\rangle\langle01|$ with coefficient $1/N$.
\item $i\ge3$: each contributes $|00\rangle\langle00|$ with coefficient $1/N$; there are $N-2$ such terms.
\end{itemize}

\paragraph{Off-diagonal contributions ($i\ne j$)\\}
The environment (qubits $3,\ldots,N$) must be all-zero in \emph{both} bra and ket. This
requires $i,j\in\{1,2\}$:
\begin{itemize}
\item $(i,j)=(1,2)$: contributes $|10\rangle\langle01|$ with coefficient $1/N$.
\item $(i,j)=(2,1)$: contributes $|01\rangle\langle10|$ with coefficient $1/N$.
\end{itemize}

Assembling in the basis $\{|00\rangle,|01\rangle,|10\rangle,|11\rangle\}$:
\begin{equation}
\rho_{12}^{W}=\frac1N
\begin{pmatrix}N-2&0&0&0\\0&1&1&0\\0&1&1&0\\0&0&0&0\end{pmatrix}
\end{equation}
X-state parameters: $a=\frac{N-2}N,\ b=c=\frac1N,\ d=0,\ z=\frac1N,\ w=0$.
Trace $=1$  .

\subsubsection{For $|\bar{W}_N\rangle$}

The density operator is
\begin{equation}
\rho_{\bar{W}_N}=|\bar{W}_N\rangle\langle\bar{W}_N|=\frac1N\sum_{i,j=1}^N
|1\cdots0_i\cdots1\rangle\langle1\cdots0_j\cdots1|.
\end{equation}

\paragraph{Diagonal contributions ($i=j$)\\}
\begin{itemize}
\item $i=1$: contributes $|01\rangle\langle01|$ with coefficient $1/N$.
\item $i=2$: contributes $|10\rangle\langle10|$ with coefficient $1/N$.
\item $i\ge3$: each contributes $|11\rangle\langle11|$ with coefficient $1/N$; there are $N-2$ such terms.
\end{itemize}

\paragraph{Off-diagonal contributions ($i\ne j$)\\}
Environment must be all-one in both bra and ket $\Rightarrow$ $i,j\in\{1,2\}$:
\begin{itemize}
\item $(i,j)=(1,2)$: contributes $|01\rangle\langle10|$ with coefficient $1/N$.
\item $(i,j)=(2,1)$: contributes $|10\rangle\langle01|$ with coefficient $1/N$.
\end{itemize}

\begin{equation}
\rho_{12}^{\bar{W}}=\frac1N
\begin{pmatrix}0&0&0&0\\0&1&1&0\\0&1&1&0\\0&0&0&N-2\end{pmatrix}
\end{equation}
X-state parameters: $a=0,\ b=c=\frac1N,\ d=\frac{N-2}N,\ z=\frac1N,\ w=0$.

\subsection{Initial Bipartite Concurrence}

Using $\mathcal{C}=2\max\{0,|z|-\sqrt{ad}\}$ (since $w=0$ for both):

\subsubsection{For $\rho_{12}^W$}
$|z|=1/N$, $\sqrt{ad}=\sqrt{(N-2)/N\cdot0}=0$:
\begin{equation}
\mathcal{C}(W_N)=\frac2N
\end{equation}

\subsubsection{For $\rho_{12}^{\bar{W}}$}
$|z|=1/N$, $\sqrt{ad}=\sqrt{0\cdot(N-2)/N}=0$:
\begin{equation}
\mathcal{C}(\bar{W}_N)=\frac2N
\end{equation}

Both states have identical initial concurrence $2/N$, decreasing inversely with system size.

\section{Noise Channels on $\rho_{12}^{W}$}
\label{app:noiseC-W}
We now specialize the general channel transformations of
Appendix~\ref{app:general} to the symmetric single-excitation state.
For $(a,b,c,d,z,w)=\left(\frac{N-2}{N},\frac1N,\frac1N,0,\frac1N,0\right),$
the vanishing $d$ element eliminates the concurrence penalty
$\sqrt{a'd'}$ whenever the channel does not populate the
$|11\rangle$ component. This observation explains the absence of
ESD under PD and AD before the explicit formulas are evaluated below.

\subsection{Phase Damping}
$z'=\frac{\sqrt{1-p}}N$, $a'=\frac{N-2}N$, $d'=0$:
\begin{equation}
\mathcal{C}^W_N(p)=\frac{2\sqrt{1-p}}N\quad\text{(no ESD).}
\end{equation}

\subsection{Amplitude Damping}
$a'=\frac{N-2+\gamma}N$, $d'=0$, $z'=\frac{\sqrt{1-\gamma}}N$:
\begin{equation}
\mathcal{C}^W_N(\gamma)=\frac{2\sqrt{1-\gamma}}N\quad\text{(no ESD).}
\end{equation}

\subsection{Depolarization}
$a'=\frac{N-2}N+\frac{2p}3\bigl(\frac1N-\frac{N-2}N\bigr)
=\frac{N-2}N-\frac{2p(N-3)}{3N}$,\;
$d'=\frac{2p}{3N}$,\;
$z'=\frac1N\bigl(1-\frac{4p}3\bigr)$.
\begin{equation}
\mathcal{C}^W_N(p)=\frac2N\max\!\left\{0,\;
\left|1-\frac{4p}{3}\right|
-\sqrt{\frac{2p(N-2)}{3}-\frac{4p^2(N-3)}{9}}\right\}
\end{equation}
Setting the argument to zero yields:
\begin{equation}
p_W^{*}=\frac3{2(N+1)}
\end{equation}

\subsection{GAD}
\label{sec:gadc-W}
$a+b=\frac{N-1}N$, $c+d=\frac1N$. Applying the general GAD transformation of Appendix~\ref{app:general}:
\begin{align}
a' &= \frac{N-2}N(1-p)+\alpha p\cdot\frac{N-1}N
    = \frac{N-2-p(N-2)+\alpha p(N-1)}{N}
    = \frac{N-2+p[\alpha(N-1)-(N-2)]}{N},\\
d' &= (1-\alpha)p\cdot\frac1N+0
    = \frac{(1-\alpha)p}{N},\\
z' &= \frac{\sqrt{1-p}}{N}.
\end{align}
Concurrence:
\begin{equation}
\mathcal{C}^W_N(p,\alpha)=\frac2N\max\!\left\{0,\;\sqrt{1-p}
-\sqrt{(1-\alpha)p\bigl[N-2+p(\alpha(N-1)-(N-2))\bigr]}\right\}
\end{equation}
The limiting cases provide useful consistency checks. At $\alpha=1$,
$d'=0$ and the result reduces to pure AD,
$\mathcal{C}=2\sqrt{1-p}/N$, with no ESD. At $\alpha=0$, one obtains
$a'=(N-2)(1-p)/N$ and $d'=p/N$, giving
\[
\mathcal{C}
=\frac{2\sqrt{1-p}}{N}
\max\{0,1-\sqrt{p(N-2)}\},
\]
with threshold $p=1/(N-2)$ when it lies inside the physical interval.

\section{Noise Channels on $\rho_{12}^{\bar{W}}$}
\label{app:noiseD-W}
The bit-flipped symmetric state occupies the complementary
$(N-1)$-excitation sector. Its reduced state is obtained from the
single-excitation reference by exchanging the corner populations,
so that $(a,b,c,d,z,w)=\left(0,\frac1N,\frac1N,\frac{N-2}{N},\frac1N,0\right).$
This exchange leaves the PD and DP results symmetric between the two
states, but becomes physically consequential under directional
amplitude damping.

\subsection{Phase Damping}
$z'=\frac{\sqrt{1-p}}N$, $a'=0$, $d'=\frac{N-2}N$:
\begin{equation}
\mathcal{C}^{\bar W}_N(p)=\frac{2\sqrt{1-p}}N\quad\text{(no ESD).}
\end{equation}

\subsection{Amplitude Damping}
$a'=\frac\gamma N$, $d'=\frac{(N-2)(1-\gamma)}N$, $z'=\frac{\sqrt{1-\gamma}}N$:
\begin{equation}
{\rho'}_{12}^{\bar W}=\frac1N\begin{pmatrix}\gamma&0&0&0\\0&1-\gamma&\sqrt{1-\gamma}&0\\0&\sqrt{1-\gamma}&1+\gamma(N-2)&0\\0&0&0&(1-\gamma)(N-2)\end{pmatrix}
\end{equation}
\begin{equation}
\mathcal{C}^{\bar W}_N(\gamma)=\frac{2\sqrt{1-\gamma}}{N}\max\!\left\{0,1-\sqrt{\gamma(N-2)}\right\},
\end{equation}
with ESD threshold
\begin{equation}
	\gamma^{*}=\frac1{N-2}\quad(N\ge4;\;\gamma^*=1\text{ at }N=3,\text{ boundary}).
\end{equation}

\subsection{Depolarization}
$a'=\frac{2p}{3N}$,\;$d'=\frac{N-2}N-\frac{2p(N-3)}{3N}$,\;
$z'=\frac1N(1-\frac{4p}3)$:
\begin{equation}
\mathcal{C}^{\bar W}_N(p)=\frac2N\max\!\left\{0,\;
\left|1-\frac{4p}{3}\right|
-\sqrt{\frac{2p(N-2)}{3}-\frac{4p^2(N-3)}{9}}\right\}
=\mathcal{C}^W_N(p)
\end{equation}
Same threshold: $p^*=\frac3{2(N+1)}$.

\subsection{GAD}
$a+b=\frac1N$, $c+d=\frac{N-1}N$:
\begin{align}
a' &= 0+\alpha p\cdot\frac1N = \frac{\alpha p}N,\\
d' &= (1-\alpha)p\cdot\frac1N+\frac{N-2}N(1-\alpha p)
    = \frac{N-2+p[1-\alpha(N-1)]}{N},\\
z' &= \frac{\sqrt{1-p}}N.
\end{align}
\begin{equation}
\mathcal{C}^{\bar W}_N(p,\alpha)=\frac2N\max\!\left\{0,\;\sqrt{1-p}
-\sqrt{\alpha p\bigl[N-2+p(1-\alpha(N-1))\bigr]}\right\}
\end{equation}
\textbf{Bit-flip symmetry:} substituting $\alpha\to1-\alpha$ in $\mathcal{C}^W_N(p,\alpha)$ gives
$\mathcal{C}^{\bar W}_N(p,\alpha)$, confirming $\mathcal{C}^W(p,\alpha)=\mathcal{C}^{\bar W}(p,1-\alpha)$.\\
At $\alpha^*=\frac12$, the two states have identical concurrence dynamics.

\section{Initial Properties of $\WL$ and $\WLb$}
\label{app:initial-VB}
This appendix derives the coherence properties and the two
inequivalent bipartite reductions of the generalized asymmetric
family. The distinction between vertex-base and base-base reductions
established here provides the structural input for the noise
calculations in Appendices~\ref{app:noiseC-VB}--\ref{app:noiseF-VB}.

We label the qubits $1,2,\ldots,N$, with qubit~1 designated as the
vertex and qubits $2,\ldots,N$ as the equivalent base qubits. All
two-qubit density matrices are written in the computational basis
$\{|00\rangle,|01\rangle,|10\rangle,|11\rangle\}$. By the permutation
symmetry of the base qubits, it is sufficient to consider
$\rho_{VB}\equiv\rho_{12}$ and $\rho_{BB}\equiv\rho_{23}$.

\subsection{The Two States}

The generalized asymmetric $N$-qubit $|W\rangle$-class state:
\begin{equation}\label{eq:WL}
\WL = \frac1{\sqrt2}\biggl|1\underbrace{0\,0\cdots0}_{N-1}\biggr\rangle
+ \frac1{\sqrt{2(N-1)}}\sum_{k=2}^N |0\cdots1_k\cdots0\rangle,
\end{equation}
which lies in the \emph{single-excitation} subspace. Its bit-flipped partner:
\begin{equation}\label{eq:WLb}
\WLb = \sigma_x^{\otimes N}\WL
= \frac1{\sqrt2}\biggl|0\underbrace{1\,1\cdots1}_{N-1}\biggr\rangle
+ \frac1{\sqrt{2(N-1)}}\sum_{k=2}^N |1\cdots0_k\cdots1\rangle,
\end{equation}
which lies in the $(N-1)$-excitation subspace.

At $N=3$: $\WL=\frac1{\sqrt2}|100\rangle+\frac12|010\rangle+\frac12|001\rangle$ is the
single-excitation bit-flipped partner of the Lohmayer state, while
$\WLb=\frac1{\sqrt2}|011\rangle+\frac12|101\rangle+\frac12|110\rangle$ \emph{is} the
two-excitation Lohmayer state used in our recent three-qubit work~\cite{bhattacharyya2026super}.

\subsection{$\ell_1$-Norm of Coherence}

Amplitudes of $\WL$: one amplitude $1/\sqrt2$ and $(N-1)$ amplitudes $1/\sqrt{2(N-1)}$.
\begin{equation}
\sum|c_i|=\frac1{\sqrt2}+(N-1)\cdot\frac1{\sqrt{2(N-1)}}
=\frac1{\sqrt2}+\sqrt{\frac{N-1}2},
\end{equation}
\begin{equation}
\Cl(\WL)=\Cl(\WLb)=\frac{N-2+2\sqrt{N-1}}2.
\end{equation}

\subsubsection{Measuring the Vertex Qubit of $\WL$}

Outcome $|1\rangle$ (prob.\ $\frac12$): residual state $|0\cdots0\rangle_{N-1}$, coherence $=0$.

Outcome $|0\rangle$ (prob.\ $\frac12$): residual state $|W_{N-1}\rangle$, coherence $=N-2$.

Average residual coherence:
\begin{equation}
\langle \Cl\rangle_V^{W^L}=\frac{N-2}2.
\end{equation}

\subsubsection{Measuring a Base Qubit ($k\ge2$) of $\WL$}

Outcome $|1\rangle$ (prob.\ $\frac1{2(N-1)}$):
residual $(N-1)$-qubit state, coherence $=0$ (product state).

Outcome $|0\rangle$ (prob.\ $\frac{2N-3}{2(N-1)}$):
residual is a renormalized asymmetric $(N-1)$-qubit state.

Average residual coherence:
\begin{equation}
\langle \Cl\rangle_B^{W^L}
=\frac{(N-2)\bigl[(N-3)+2\sqrt{N-1}\bigr]}{2(N-1)}.
\end{equation}

\subsubsection{For $\WLb$}
By global bit-flip symmetry, all average residual coherences are identical to those of $\WL$.

\subsection{Bipartite Reduced Density Matrices of $\WL$}

\subsubsection{Vertex-Base: $\rho_{12}^{W^L}$}

From $\WL$, tracing out qubits $3,\ldots,N$:

\paragraph{Diagonal contributions\\}
\begin{itemize}
\item $|1\,0\cdots0\rangle$ ($k=1$ term): contributes $|10\rangle\langle10|$ with coefficient $\frac12$.
\item $|0\cdots1_2\cdots0\rangle$ ($k=2$ term): contributes $|01\rangle\langle01|$ with coefficient $\frac1{2(N-1)}$.
\item $k\ge3$: each contributes $|00\rangle\langle00|$ with coefficient $\frac1{2(N-1)}$;
      there are $N-2$ such terms, giving total $\frac{N-2}{2(N-1)}$.
\end{itemize}

\paragraph{Off-diagonal contributions\\}
Only the $k=1$ and $k=2$ terms share the same environment state (all zeros for qubits $3,\ldots,N$):
\begin{itemize}
\item Cross-term: $\frac1{\sqrt2}\cdot\frac1{\sqrt{2(N-1)}}=\frac1{2\sqrt{N-1}}$.
\end{itemize}

\begin{equation}\label{eq:rhoVB-WL}
\rho_{12}^{W^L}=
\begin{pmatrix}
\frac{N-2}{2(N-1)}&0&0&0\\[4pt]
0&\frac1{2(N-1)}&\frac1{2\sqrt{N-1}}&0\\[4pt]
0&\frac1{2\sqrt{N-1}}&\frac12&0\\[4pt]
0&0&0&0
\end{pmatrix}
\end{equation}
Parameters: $a=\frac{N-2}{2(N-1)},\;b=\frac1{2(N-1)},\;c=\frac12,\;d=0,\;z=\frac1{2\sqrt{N-1}},\;w=0$.

\subsubsection{Base-Base: $\rho_{23}^{W^L}$}

Tracing out qubits $1,4,5,\ldots,N$:

\paragraph{Diagonal contributions\\}
\begin{itemize}
\item $k=1$ (vertex excited): contributes $|00\rangle\langle00|$ with coefficient $\frac12$.
\item $k=2$: contributes $|10\rangle\langle10|$ with coefficient $\frac1{2(N-1)}$.
\item $k=3$: contributes $|01\rangle\langle01|$ with coefficient $\frac1{2(N-1)}$.
\item $k\ge4$: each contributes $|00\rangle\langle00|$ with coefficient $\frac1{2(N-1)}$;
      there are $N-3$ such terms, total $\frac{N-3}{2(N-1)}$.
\end{itemize}
Total $|00\rangle$ population: $\frac12+\frac{N-3}{2(N-1)}=\frac{N-2}{N-1}$.

\paragraph{Off-diagonal contributions\\}
Only $k=2,3$ share the same environment (qubit~1 in $|0\rangle$, qubits $4,\ldots,N$ all $|0\rangle$):
coefficient $\frac1{2(N-1)}$.

\begin{equation}
\rho_{23}^{W^L}=
\begin{pmatrix}
\frac{N-2}{N-1}&0&0&0\\[4pt]
0&\frac1{2(N-1)}&\frac1{2(N-1)}&0\\[4pt]
0&\frac1{2(N-1)}&\frac1{2(N-1)}&0\\[4pt]
0&0&0&0
\end{pmatrix}
\end{equation}
Parameters: $a=\frac{N-2}{N-1},\;b=c=\frac1{2(N-1)},\;d=0,\;z=\frac1{2(N-1)},\;w=0$.

\subsection{Bipartite Reduced Density Matrices of $\WLb$, via Bit-Flip Symmetry}

Under global bit-flip ($\sigma_x^{\otimes N}$), the two-qubit reduced state transforms as
$(\sigma_x\otimes\sigma_x)\rho(\sigma_x\otimes\sigma_x)$, which swaps $a\leftrightarrow d$ and
$b\leftrightarrow c$ while preserving $z$ and $w$:

\begin{equation}
\rho_{12}^{\overline{W^L}}=
\begin{pmatrix}
0&0&0&0\\[4pt]
0&\frac12&\frac1{2\sqrt{N-1}}&0\\[4pt]
0&\frac1{2\sqrt{N-1}}&\frac1{2(N-1)}&0\\[4pt]
0&0&0&\frac{N-2}{2(N-1)}
\end{pmatrix}
\quad\text{(VB)}
\end{equation}

\begin{equation}
\rho_{23}^{\overline{W^L}}=
\begin{pmatrix}
0&0&0&0\\[4pt]
0&\frac1{2(N-1)}&\frac1{2(N-1)}&0\\[4pt]
0&\frac1{2(N-1)}&\frac1{2(N-1)}&0\\[4pt]
0&0&0&\frac{N-2}{N-1}
\end{pmatrix}
\quad\text{(BB)}
\end{equation}

\subsection{Initial Bipartite Concurrence}

\begin{align}
\mathcal{C}_{VB}(W^L) &= 2\cdot\frac1{2\sqrt{N-1}} = \frac1{\sqrt{N-1}} = \mathcal{C}_{VB}(\overline{W^L}),\\
\mathcal{C}_{BB}(W^L) &= 2\cdot\frac1{2(N-1)} = \frac1{N-1} = \mathcal{C}_{BB}(\overline{W^L}).
\end{align}

\paragraph{Concurrence Hierarchy\\}
For all $N\ge3$:
\begin{equation}
\mathcal{C}_{VB}=\frac1{\sqrt{N-1}} > \mathcal{C}_W=\frac2N > \mathcal{C}_{BB}=\frac1{N-1}
\end{equation}

\section{Noise on $\rho_{12}^{W^L}$ (Vertex-Base) }
\label{app:noiseC-VB}
We first consider the vertex-base reduction of the asymmetric
single-excitation state. Substituting $(a,b,c,d,z,w)=
\left(
\frac{N-2}{2(N-1)},\frac1{2(N-1)},\frac12,0,
\frac1{2\sqrt{N-1}},0
\right)
$
into Appendix~\ref{app:general} yields the channel-dependent
expressions below. The vanishing initial $d$ element is responsible
for the ESD-free PD and AD dynamics of this pair class.

\subsection{Phase Damping}
$z'=\frac{\sqrt{1-p}}{2\sqrt{N-1}}$, $d'=0$:
\begin{equation}
\mathcal{C}_{VB}^{W^L}(p)=\frac{\sqrt{1-p}}{\sqrt{N-1}}=\mathcal{C}_{VB}(0)\sqrt{1-p}\quad\text{(no ESD).}
\end{equation}

\subsection{Amplitude Damping}
$a'=\frac{N-2+\gamma}{2(N-1)}$, $d'=0$, $z'=\frac{\sqrt{1-\gamma}}{2\sqrt{N-1}}$:
\begin{equation}
\mathcal{C}_{VB}^{W^L}(\gamma)=\frac{\sqrt{1-\gamma}}{\sqrt{N-1}}\quad\text{(no ESD for any }N).
\end{equation}

\subsection{Depolarization}
The transformed matrix elements relevant to the concurrence are
$a'=\frac{3(N-2)-2p(N-3)}{6(N-1)}$,
$d'=\frac{p}{3}$, and $z'=\frac{1}{2\sqrt{N-1}}\left(1-\frac{4p}{3}\right)$.

\begin{equation}
\mathcal{C}_{VB}^{W^L}(p)=\frac1{\sqrt{N-1}}\max\!\left\{0,\;\left|1-\frac{4p}{3}\right|
-\sqrt{\frac{2p[3(N-2)-2p(N-3)]}{9}}\right\}
\end{equation}
Setting the argument to zero: the threshold quadratic $4(N+1)p^2-6(N+2)p+9=0$ gives
\begin{equation}
p_{VB}^{*}=\frac3{2(N+1)}
\end{equation}

\subsection{GAD}
\label{sec:gadc-VB-WL}
Using the GAD transformation of Appendix~\ref{app:general} with
$a+b=\frac12$:
\begin{align}
a' &= \frac{N-2}{2(N-1)}(1-p)+\alpha p\cdot\frac12
    = \frac{(N-2)(1-p)+\alpha p(N-1)}{2(N-1)}
    = \frac{N-2+p[\alpha(N-1)-(N-2)]}{2(N-1)},\\
d' &= (1-\alpha)p\cdot\frac12 = \frac{(1-\alpha)p}2,\\
z' &= \frac{\sqrt{1-p}}{2\sqrt{N-1}}.
\end{align}
The ESD product:
\begin{equation}
a'd' = \frac{(1-\alpha)p\bigl[N-2+p(\alpha(N-1)-(N-2))\bigr]}{4(N-1)}.
\end{equation}
Concurrence:
\begin{equation}
\mathcal{C}_{VB}^{W^L}(p,\alpha)=\frac1{\sqrt{N-1}}\max\!\left\{0,\;\sqrt{1-p}
-\sqrt{(1-\alpha)p\bigl[N-2+p(\alpha(N-1)-(N-2))\bigr]}\right\}
\end{equation}

The limiting cases provide useful consistency checks. At $\alpha=1$,
$d'=0$, and the result reduces to
\[
C=\frac{\sqrt{1-p}}{\sqrt{N-1}},
\]
with no ESD. At $\alpha=0$,
$a'=(N-2)(1-p)/[2(N-1)]$ and $d'=p/2$, giving
\[
C=\frac{\sqrt{1-p}}{\sqrt{N-1}}
\max\{0,1-\sqrt{p(N-2)}\}.
\]
The corresponding threshold is $p=1/(N-2)$ when it lies inside the
physical interval. At $N=3$, the complementary-sector result derived
below reproduces the corresponding expression of our recent
three-qubit work~\cite{bhattacharyya2026super}.

\section{Noise on $\rho_{23}^{W^L}$ (Base-Base)}
\label{app:noiseD-VB}
The base-base reduction belongs to the same single-excitation
sector but has a different population structure and a smaller
initial coherence than the vertex-base reduction. Using
$
(a,b,c,d,z,w)=
\left(
\frac{N-2}{N-1},\frac1{2(N-1)},\frac1{2(N-1)},0,
\frac1{2(N-1)},0
\right),
$
we evaluate the same four channels in order to isolate the effect of
pair-class structure at fixed excitation sector.

\subsection{Phase Damping}
$z'=\tfrac{\sqrt{1-p}}{2(N-1)}$, $d'=0$:
\begin{equation}
\mathcal{C}_{BB}^{W^L}(p)=\frac{\sqrt{1-p}}{N-1}=\mathcal{C}_{BB}(0)\sqrt{1-p}\quad\text{(no ESD).}
\end{equation}

\subsection{Amplitude Damping}
$a'=\tfrac{2(N-2)+\gamma}{2(N-1)}$, $d'=0$, $z'=\tfrac{\sqrt{1-\gamma}}{2(N-1)}$:
\begin{equation}
\mathcal{C}_{BB}^{W^L}(\gamma)=\frac{\sqrt{1-\gamma}}{N-1}\quad\text{(no ESD for any }N).
\end{equation}

\subsection{Depolarization}
$a'=\frac{N-2}{N-1}+\frac{2p}{3}\bigl(\frac1{2(N-1)}-\frac{N-2}{N-1}\bigr)
=\frac{3(N-2)+p(5-2N)}{3(N-1)}$,\\[2pt]
$d'=\frac{2p}{3}\cdot\frac1{2(N-1)}=\frac p{3(N-1)}$,\\[2pt]
$z'=\frac1{2(N-1)}\bigl(1-\frac{4p}3\bigr)$.

\begin{equation}
\mathcal{C}_{BB}^{W^L}(p)
=\frac1{N-1}\max\!\left\{0,\;
\left|1-\frac{4p}{3}\right|
-\frac23\sqrt{p[3(N-2)+p(5-2N)]}\right\}.
\end{equation}
Setting the argument to zero: $4(2N-1)p^2-12Np+9=0$, discriminant $144(N-1)^2$ (a perfect
square), giving
\begin{equation}
p_{BB}^{*}=\frac{3}{2(2N-1)}.
\end{equation}

\subsection{GAD}
\label{sec:gadc-BB-WL}
$a+b=\frac{N-2}{N-1}+\frac1{2(N-1)}=\frac{2N-3}{2(N-1)}$:
\begin{align}
a' &= \frac{N-2}{N-1}(1-p)+\alpha p\cdot\frac{2N-3}{2(N-1)}
    = \frac{2(N-2)(1-p)+\alpha p(2N-3)}{2(N-1)},\\
d' &= (1-\alpha)p\cdot\frac1{2(N-1)} = \frac{(1-\alpha)p}{2(N-1)},\\
z' &= \frac{\sqrt{1-p}}{2(N-1)}.
\end{align}

\begin{equation}
\mathcal{C}_{BB}^{W^L}(p,\alpha)=\frac1{N-1}\max\!\left\{0,\;\sqrt{1-p}
-\sqrt{(1-\alpha)p\bigl[2(N-2)(1-p)+\alpha p(2N-3)\bigr]}\right\}.
\end{equation}

The limiting cases again provide a direct check of the result. At
$\alpha=1$, $d'=0$, and the concurrence reduces to
\[
C=\frac{\sqrt{1-p}}{N-1},
\]
with no ESD. At $\alpha=0$, one obtains
\[
C=\frac{\sqrt{1-p}}{N-1}
\max\{0,1-\sqrt{2p(N-2)}\},
\]
with threshold $p=1/[2(N-2)]$.

\section{Noise on $\rho_{12}^{\overline{W^L}}$ (Vertex-Base)}
\label{app:noiseE-VB}
We next consider the vertex-base reduction in the complementary
$(N-1)$-excitation sector. Global bit-flip symmetry exchanges the
corner populations of the corresponding $W^L$ reduction, while
preserving the magnitude of the coherence. This relation is
sufficient to anticipate identical PD and DP dynamics but different
AD behaviour between the two excitation sectors. Hence, we substitute $(a,b,c,d,z,w)=\bigl(0,\tfrac12,\tfrac1{2(N-1)},\tfrac{N-2}{2(N-1)},\tfrac1{2\sqrt{N-1}},0\bigr)$
into Appendix~\ref{app:general}.

\subsection{Phase Damping}
$a'=0$, $z'=\tfrac{\sqrt{1-p}}{2\sqrt{N-1}}$:
\begin{equation}
\mathcal{C}_{VB}^{\overline{W^L}}(p)=\frac{\sqrt{1-p}}{\sqrt{N-1}}=\mathcal{C}_{VB}^{W^L}(p),\quad\text{(no ESD).}
\end{equation}

\subsection{Amplitude Damping}
$a'=\tfrac\gamma2$, $d'=\tfrac{(1-\gamma)(N-2)}{2(N-1)}$, $z'=\tfrac{\sqrt{1-\gamma}}{2\sqrt{N-1}}$:
\begin{equation}
\mathcal{C}_{VB}^{\overline{W^L}}(\gamma)=\frac{\sqrt{1-\gamma}}{\sqrt{N-1}}\max\!\left\{0,1-\sqrt{\gamma(N-2)}\right\}.
\end{equation}
\begin{equation}
\gamma_{VB}^{*}=\frac1{N-2}.
\end{equation}
(At $N=3$, the formal solution is $\gamma_{VB}=1$, so no ESD occurs;
$\gamma_{VB}^{*}<1$ and represents an ESD threshold for $N\ge4$.)

\subsection{Depolarization}
$a'=\tfrac p3$, $d'=\tfrac{N-2}{2(N-1)}-\tfrac{p(N-3)}{3(N-1)}$,
$z'=\tfrac1{2\sqrt{N-1}}\bigl(1-\tfrac{4p}3\bigr)$ --- identical $a'd'$ product to
Appendix~\ref{app:noiseC-VB} (only $a'\leftrightarrow d'$ swap, which doesn't change the product):
\begin{equation}
\mathcal{C}_{VB}^{\overline{W^L}}(p)=\mathcal{C}_{VB}^{W^L}(p),\qquad p_{VB}^{*}=\frac3{2(N+1)}
\end{equation}

$\sqrt{a'd'}$ is invariant under $a\leftrightarrow d$, hence $\mathcal{C}_{VB}^{\overline{W^L}}=\mathcal{C}_{VB}^{W^L}$.

\subsection{GAD}
\label{sec:gadc-VB-WLbar}
$a+b=\frac12$, $c+d=\frac12$:
\begin{align}
a' &= 0+\alpha p\cdot\frac12 = \frac{\alpha p}2,\\
d' &= (1-\alpha)p\cdot\frac1{2(N-1)}+\frac{N-2}{2(N-1)}(1-\alpha p)
    = \frac{N-2+p[1-\alpha(N-1)]}{2(N-1)},\\
z' &= \frac{\sqrt{1-p}}{2\sqrt{N-1}}.
\end{align}
\begin{equation}
\mathcal{C}_{VB}^{\overline{W^L}}(p,\alpha)=\frac1{\sqrt{N-1}}\max\!\left\{0,\;\sqrt{1-p}
-\sqrt{\alpha p\bigl[N-2+p(1-\alpha(N-1))\bigr]}\right\}
\end{equation}

\textbf{Verification at $N=3$:}
$a'=\alpha p/2$, $d'=(1+p-2\alpha p)/4$, $z'=\sqrt{(1-p)/8}$.
Concurrence $=(1/\sqrt2)\max\{0,\sqrt{1-p}-\sqrt{\alpha p(1+p-2\alpha p)}\}$.
This exactly reproduces the corresponding three-qubit result of
Ref.~\cite{bhattacharyya2026super}.

\section{Noise on $\rho_{23}^{\overline{W^L}}$ (Base-Base)}
\label{app:noiseF-VB}
Finally, we evaluate the base-base reduction in the
$(N-1)$-excitation sector. This case combines the peripheral
base-base geometry with the excitation sector that is sensitive to
amplitude damping, and therefore provides the comparison needed to
identify the lower BB ESD threshold as a same-sector structural
effect. Accordingly, we substitute $(a,b,c,d,z,w)=\bigl(0,\tfrac1{2(N-1)},\tfrac1{2(N-1)},\tfrac{N-2}{N-1},\tfrac1{2(N-1)},0\bigr)$
into Appendix~\ref{app:general}.

\subsection{Phase Damping}
\begin{equation}
\mathcal{C}_{BB}^{\overline{W^L}}(p)=\frac{\sqrt{1-p}}{N-1}=\mathcal{C}_{BB}^{W^L}(p)\quad\text{(no ESD).}
\end{equation}

\subsection{Amplitude Damping}
$a'=\tfrac\gamma{2(N-1)}$, $d'=\tfrac{(1-\gamma)(N-2)}{N-1}$, $z'=\tfrac{\sqrt{1-\gamma}}{2(N-1)}$:
\begin{equation}
\mathcal{C}_{BB}^{\overline{W^L}}(\gamma)=\frac{\sqrt{1-\gamma}}{N-1}\max\!\left\{0,1-\sqrt{2\gamma(N-2)}\right\},
\end{equation}
\begin{equation}
\gamma_{BB}^{*}=\frac1{2(N-2)}.
\end{equation}

$\gamma_{BB}^*=\tfrac12\gamma_{VB}^*$ for all $N\ge3$: the BB pair's
$\rho_{44}=\tfrac{N-2}{N-1}$ is twice the VB pair's
$\tfrac{N-2}{2(N-1)}$, halving the threshold. At $N=3$,
$\gamma_{VB}^*=1$ lies at the physical boundary, while
$\gamma_{BB}^*=1/2$ corresponds to ESD.

\subsection{Depolarization}
$a'=\tfrac{p}{3(N-1)}$, $d'=\tfrac{N-2}{N-1}-\tfrac{p(2N-5)}{3(N-1)}
=\tfrac{3(N-2)-p(2N-5)}{3(N-1)}$,
$z'=\tfrac1{2(N-1)}\bigl(1-\tfrac{4p}3\bigr)$:
\begin{equation}
\mathcal{C}_{BB}^{\overline{W^L}}(p)
=\frac1{N-1}\max\!\left\{0,\;
\left|1-\frac{4p}{3}\right|
-\frac23\sqrt{p[3(N-2)-p(2N-5)]}\right\}.
\end{equation}
The threshold quadratic is identical to Appendix~\ref{app:noiseD-VB}'s ($4(2N-1)p^2-12Np+9=0$):
\begin{equation}
p_{BB}^{*}=\frac3{2(2N-1)}=p_{BB}^{*,W^L}
\end{equation}

\subsection{GAD}
\label{sec:gadc-BB-WLbar}
$a+b=\frac1{2(N-1)}$, $c+d=\frac{2N-3}{2(N-1)}$:
\begin{align}
a' &= 0+\alpha p\cdot\frac1{2(N-1)} = \frac{\alpha p}{2(N-1)},\\
d' &= (1-\alpha)p\cdot\frac1{2(N-1)}+\frac{N-2}{N-1}(1-\alpha p)
    = \frac{2(N-2)+p[1-\alpha(2N-3)]}{2(N-1)},\\
z' &= \frac{\sqrt{1-p}}{2(N-1)}.
\end{align}
\begin{equation}
\mathcal{C}_{BB}^{\overline{W^L}}(p,\alpha)=\frac1{N-1}\max\!\left\{0,\;\sqrt{1-p}
-\sqrt{\alpha p\bigl[2(N-2)+p(1-\alpha(2N-3))\bigr]}\right\}
\end{equation}

\textbf{Verification against our recent three-qubit work at $N=3$:}
$a'=\alpha p/4$, $d'=(2+p-3\alpha p)/4$, $z'=\sqrt{1-p}/4$.
Concurrence $=(1/2)\max\{0,\sqrt{1-p}-\sqrt{\alpha p(2+p-3\alpha p)}\}$.
This exactly reproduces the corresponding three-qubit result of Ref.~\cite{bhattacharyya2026super}. 

\subsection{Cross-State Symmetry and the $\alpha=\tfrac{1}{2}$ Symmetry Point}

Substituting $\alpha\to1-\alpha$ in the ESD products of $\mathcal{C}_{VB}^{W^L}$ and $\mathcal{C}_{BB}^{W^L}$
(Appendices~\ref{app:noiseC-VB}, \ref{app:noiseD-VB}) reproduces exactly the ESD products of
$\mathcal{C}_{VB}^{\overline{W^L}}$ and $\mathcal{C}_{BB}^{\overline{W^L}}$ above:
\begin{equation}
\mathcal{C}_{VB}^{W^L}(p,\alpha)=\mathcal{C}_{VB}^{\overline{W^L}}(p,1-\alpha),\qquad
\mathcal{C}_{BB}^{W^L}(p,\alpha)=\mathcal{C}_{BB}^{\overline{W^L}}(p,1-\alpha).
\end{equation}

\paragraph{Proof sketch for VB\\}
The penalty term of $\mathcal{C}_{VB}^{W^L}$ is $(1-\alpha)p[N-2+p(\alpha(N-1)-(N-2))]$.
Under $\alpha\to1-\alpha$:
\begin{align}
&\alpha\,p\bigl[N-2+p\bigl((1-\alpha)(N-1)-(N-2)\bigr)\bigr]\\
&=\alpha\,p\bigl[N-2+p\bigl(1-\alpha(N-1)\bigr)\bigr],
\end{align}
which is exactly the penalty term of $\mathcal{C}_{VB}^{\overline{W^L}}$. The coherence prefactor $\sqrt{1-p}$
is $\alpha$-independent, so it is unaffected.  \\

At $\alpha^{*}=\frac12$, the two states have identical concurrence
dynamics within each corresponding pair class, VB and BB, for all
$N\ge3$ and all $p\in[0,1]$.\\

\paragraph{Limits\\}
At $\alpha=1$ (pure damping), the pairs of $\WL$ have no ESD.
For $\WLb$, the BB pair undergoes ESD at
$\gamma_{BB}^{*}=1/[2(N-2)]$ for all $N\ge3$, while the VB pair
undergoes ESD at $\gamma_{VB}^{*}=1/(N-2)$ for $N\ge4$; at $N=3$,
the VB concurrence vanishes only at the boundary $\gamma=1$.
At $\alpha=0$ (pure excitation), the roles of $\WL$ and $\WLb$
interchange.

Taken together, Appendices~\ref{app:noiseC-VB}--\ref{app:noiseF-VB}
show explicitly that the bit-flip relation between the two excitation
sectors is preserved by PD and DP and appears under GAD as the exact
covariance $\alpha\leftrightarrow1-\alpha$, whereas AD distinguishes
the sectors through their different population structure. These
derivations provide the algebraic basis for the same-sector and
cross-sector comparisons discussed in Sections~\ref{sec:asym-dyn}
and~\ref{sec:reinterp}.


%

\end{document}